\documentclass[12pt]{article}
\pdfoutput=1
\usepackage{tikz}
\usetikzlibrary{matrix,arrows,decorations.pathmorphing,decorations.pathreplacing}
\usepackage{jheppub}
\usepackage{amsmath,amssymb,euscript,array,mathrsfs,appendix,ctable,marvosym}
\usepackage{arydshln}
\usepackage{todonotes}
\usepackage{graphicx}
\usepackage[normalem]{ulem}

\input{tcilatex}

\newcommand{\EQ}[1]{\begin{equation}\begin{split} #1
\end{split}\end{equation}}

\title{Exploring The Lambda Model Of The Hybrid Superstring}
\author{David M. Schmidtt}

\affiliation{Instituto de F\'\i sica Te\'orica IFT/UNESP, Rua Dr. Bento Teobaldo Ferraz 271, Bloco II, CEP 01140-070, S\~ao Paulo-SP, Brasil}

\emailAdd{david.schmidtt@gmail.com}

\abstract{The purpose of this contribution is to initiate the study of integrable
deformations for different superstring theory formalisms that manifest the property of
(classical) integrability. In this paper we choose the hybrid formalism of the superstring in
the background $AdS_{2}\times S^{2}$ and explore in detail the most immediate
consequences of its $\lambda$-deformation. The resulting action functional corresponds to 
the $\lambda $-model of the matter part of the fairly more sophisticated pure spinor 
formalism, which is also known to be classical integrable. In particular, the deformation preserves
the integrability and the one-loop conformal invariance of its parent theory, hence being a marginal deformation.}

\begin{document}

\maketitle

\newpage

\section{Introduction}

During the last few years, two novel integrable deformations of string and
superstring $\sigma $-models have received a considerable great deal of
attention due, in part, to their potential applications to the $AdS/CFT$
correspondence. The first kind of integrable field theory is known under the
name of the $\eta $-deformation and leads to a quantum group $q$-deformation
of its parent $\sigma $-model S-matrix with a real parameter $q\in 
%TCIMACRO{\U{211d} }%
%BeginExpansion
\mathbb{R}
%EndExpansion
$. For $\sigma $-models on the bosonic cosets $F/G$ and for the Green-Schwarz (GS)
superstring $\sigma $-model on the background $AdS_{5}\times S^{5}$, their respective
deformations were presented in \cite{eta bos,eta ads5xs5,ded eta ads5xs5} as
natural generalizations of the Yang-Baxter type deformation of the principal
chiral model originally constructed in \cite{Klimcik}. The second kind of integrable
field theory is known as the $\lambda $-deformation\footnote{%
Or $k$-deformation. We will bow, however, to the more common name of $%
\lambda $-deformation (or $\lambda $-model) despite of the fact that the
true quantum deformation parameter depends on the WZW level $k\in 
%TCIMACRO{\U{2124} }%
%BeginExpansion
\mathbb{Z}
%EndExpansion
$ and not on the Lagrangian deformation parameter $\lambda .$} and basically
leads to a quantum group $q$-deformation of its parent $\sigma $-model
S-matrix \cite{S1,S2,S3} this time with a root-of-unity parameter $q^{N}=1$, for some $N\in 
%TCIMACRO{\U{2124} }%
%BeginExpansion
\mathbb{Z}
%EndExpansion
.$ For the $\sigma $-models on the bosonic cosets $F/G$ and for the $AdS_{5}\times
S^{5}$ GS superstring $\sigma $-model the corresponding
deformations were introduced in \cite{k-def bos,k-def fer} as
generalizations of the deformation of the Non-Abelian T-dual of the
principal chiral model initially constructed in \cite{Sfetsos}. In this paper we will refer to these two kinds of
deformed $\sigma $-models as $\eta $-models and $\lambda $-models. These two types of
integrable field theories does not seem to be related at first sight or
to have something in common as they have very different action functionals and properties
but, remarkably, it turns out to be that under certain circumstances they form a 
sigma model pair (at least classically) under the so-called Poisson-Lie T-duality\footnote{This classical equivalence might have an interesting physical 
interpretation if realized on the dual gauge theory side under the light of the AdS/CFT duality.} \cite{relation,Hoare 1,R matrices,
Klimcik 1, Klimcik 2}. 
Recently, the properties of both approaches were combined into the so-called generalized $\lambda$-deformations \cite{generalized}, which is the largest family of string theory integrable deformations known to date. This larger theory have been considered in more detail in \cite{yuri}.

Despite of the fact that both deformations preserve the very stringent
property of integrability present in their parent $\sigma $-models, in the
superstring theory context there is a side requirement which is extremely
important and that must hold if the geometry associated to the $\eta $%
-model or the $\lambda $-model is to be considered as a consistent
deformed string theory background, i.e, a physical deformation. Thence of some relevance to 
quantum superstring theory and to the $AdS/CFT$
correspondence. This extra requirement is that the background fields (metric, dilaton, RR
fluxes, B-field etc) of the deformed theory must organize themselves into a
solution of an associated set of supergravity equations of motion.
Unfortunately, the $\eta $-model of the $AdS_{5}\times S^{5}$ GS superstring
does not surpasses this test \cite{fail} but a milder version of it instead \cite{ads5xs5}. Fortunately, the $\lambda $-models associated to the GS 
superstrings in the backgrounds $AdS_{n}\times S^{n}, n=2,3,5$ do as have been recently shown, respectively, in \cite{lambda back,lambda ads3xs3,ads5xs5}. 
For previous treatments see \cite{spacetimes,Demulder}. These results are very encouraging and
favors the $\lambda $-deformations in this respect but also raises a very interesting
question: is this result unique to the GS formalism? or does it extends to
other approaches to superstring theory as well?. After all, we know that
there are also available in the literature the Ramond-Neveu-Schwarz (RNS), the hybrid and the pure
spinor (PS) formalism, just to mention the most common formulations, which can be used to suit different needs and purposes.

Due to its inherent simplicity when compared to other approaches, in this paper
we choose the hybrid formulation \citep{Hybrid} of superstring theoy on $AdS_{2}\times
S^{2}$ to initiate the study of this question (although the supercoset $AdS_{3}\times S^{3}$
can be treated along the same lines too \cite{hybrid talk}). The hybrid formalism is a
crossbreeding between the RNS and the GS formalisms combining the
advantages of both approaches. For instance, it uses space-time spinors allowing the introduction of
RR fields like in the GS approach but in flat space it reduces to a free theory like in the RNS approach,
so quantization is straightforward. A covariant quantization preserving manifest space-time 
supersymmetry is also possible dispensing the use of the
light-cone gauge prevalent to the GS approach. The kappa symmetry proper of the GS superstring is replaced by a world-sheet superconformal invariance related to a BRST symmetry that is used to remove unphysical degrees of freedom etc. Of course, despite of its similitude with the GS formalism (on some respects) it provides a very different approach 
for treating the superstring. For further details on the properties and applications of this formalism, 
see for example \cite{Hybrid,Nathan,hybrid talk,NVW,Swieca,U(5),Covariant CY} and references therein.

The basic goal of this paper is to explore the most direct consequences of the $\lambda$-deformation for 
this simpler case keeping in mind the $AdS_{5}\times S^{5}$ supercoset for a future work 
as it requires instead the use of the pure spinor formalism \cite{PS,ICTP}, which is fairly more complex and
where, as we will see below, the introduction of the deformation is more delicate than in the present situation.
One of the goals of working out this kind of problem is to further elucidate
and understand the very structure of the $\lambda$-deformation itself from an integrable field theory point
of view by testing it on different scenarios.

The paper is organized as follows. In section \ref{s2}, we introduced the $\lambda$-model for the hybrid
superstring on $AdS_{2}\times S^{2}$ and display its properties, there we disregard the compactification manifold in order to keep the discussion simpler. In section \ref{s3}, we pursue Dirac's 
procedure and identify the phase space constraints. In section \ref{s4}, we prove the classical integrability of the
deformed hybrid formalism and show that it has the same integrable structure
than the $\lambda$-model of the Green-Schwarz formalism. In section \ref{s5}, we
elaborate on the $\lambda \rightarrow 0$ limit and show how the deformed
theory presents a light-cone splitting on its current algebra revealing a 2d Lorentz invariance suggesting that the
theory might be simpler in this limit. In section \ref{N=2}, we provide evidence that the conjectured
$N=(2,2)$ superconformal invariance of the undeformed hybrid superstring might be extended also to its 
$\lambda$-deformed partner. In section \ref{s7}, we show that the $\lambda $-deformation is
marginal to one-loop in $1/k$ but exact in the deforming parameter $\lambda $, hence preserving the
one-loop conformal invariance of the original theory. In section \ref{s8} we speculate on a possible deformation of the PS formalism
and point out a subtlety of the deformation in the current form.
In section \ref{conclusions}, we write some
concluding remarks and comment on some possible future directions of research. There are two appendices.

\section{The $\protect\lambda $-model of the hybrid superstring on $%
AdS_{2}\times S^{2}$}\label{s2}

Start by introducing some basic information\footnote{%
The 2d notation used in this paper is: $x^{\pm }=t\pm x,$ $\partial
_{\pm }=\frac{1}{2}(\partial _{0}\pm \partial _{1}),$ $\eta _{\mu \nu
}=diag(1,-1)$ and $\epsilon ^{01}=1.$ We also have that $a_{\pm }=\frac{1}{2}%
(a_{0}\pm a_{1}).$}. Consider the Lie superalgebra $\mathfrak{f=psu(}1,1|2%
\mathfrak{)}$ and its $%
%TCIMACRO{\U{2124} }%
%BeginExpansion
\mathbb{Z}
%EndExpansion
_{4}$ decomposition induced by the automorphism $\Phi $%
\begin{equation}
\Phi (\mathfrak{f}^{(m)})=i^{m}\mathfrak{f}^{(m)},\text{ \ \ }\mathfrak{f=}%
\bigoplus\nolimits_{i=0}^{3}\mathfrak{f}^{(i)},\text{ \ \ }[\mathfrak{f}%
^{(m)},\mathfrak{f}^{(n)}]\subset \mathfrak{f}^{(m+n)\func{mod}4},\text{ \ \ 
}  \label{auto}
\end{equation}%
where $m,n=0,1,2,3$. From this decomposition we associate the usual twisted loop superagebra%
\begin{equation}
\widehat{\mathfrak{f}}=\bigoplus\nolimits_{n\in 
%TCIMACRO{\U{2124} }%
%BeginExpansion
\mathbb{Z}
%EndExpansion
}\left( \bigoplus\nolimits_{i=0}^{3}\mathfrak{f}^{(i)}\otimes
z^{4n+i}\right) =\bigoplus\nolimits_{n\in 
%TCIMACRO{\U{2124} }%
%BeginExpansion
\mathbb{Z}
%EndExpansion
}\widehat{\mathfrak{f}}^{(n)}.  \label{loop superalgebra}
\end{equation}

The action functional for the $\lambda $-models is given by the general expression
\begin{equation}
S_{\lambda }=S_{F/F}(\mathcal{F},A_{\mu })-\frac{k}{\pi }\dint_{\Sigma
}d^{2}x\left\langle A_{+}(\Omega -1)A_{-}\right\rangle ,\text{ \ \ }k\in 
%TCIMACRO{\U{2124} }%
%BeginExpansion
\mathbb{Z}
%EndExpansion
,  \label{k-deformed action}
\end{equation}%
where $\left\langle \ast ,\ast \right\rangle =Str(\ast ,\ast )$ is the
supertrace in some faithful representation of $\mathfrak{f}$, $\Sigma =%
%TCIMACRO{\U{211d} }%
%BeginExpansion
\mathbb{R}
%EndExpansion
\times S^{1}$ is the world-sheet manifold with the topology of a closed string 
(a cylinder) and%
\begin{equation}
\Omega =P^{(0)}+\lambda ^{-3}P^{(1)}+\lambda ^{-2}P^{(2)}+\lambda
^{-1}P^{(3)},\text{ \ \ }\lambda ^{-2}=1+\kappa ^{2}/k  \label{projector}
\end{equation}%
is the omega projector that defines the $\lambda $-deformed hybrid superstring. It is worth highlighting the difference with
the $\Omega$ projector of the $\lambda$-deformed GS superstring \cite{k-def fer}
\begin{equation}
\Omega =P^{(0)}+\lambda ^{-1}P^{(1)}+\lambda ^{-2}P^{(2)}+\lambda
P^{(3)}.
\end{equation}
The main difference is that while the former introduce a kinetic term for the current components along fermionic coset directions, the latter forbids such a term and this crucial difference has important consequences for the symmetry structure of both theories and also for their quantization. More on this below.

Above,
we have that%
\begin{equation}
S_{F/F}(\mathcal{F},A_{\mu })=S_{WZW}(\mathcal{F})-\frac{k}{\pi }%
\dint_{\Sigma }d^{2}x\left\langle A_{+}\partial _{-}\mathcal{FF}^{-1}-A_{-}%
\mathcal{F}^{-1}\partial _{+}\mathcal{F-}A_{+}\mathcal{F}A_{-}\mathcal{F}%
^{-1}+A_{+}A_{-}\right\rangle ,
\end{equation}%
where $S_{WZW}(\mathcal{F})$ is the usual WZW model action. Note that the gauge field $A_{\pm}\in\mathfrak{f}$ takes values on the whole Lie superalgebra. The
action \eqref{k-deformed action} is universal and each $\lambda$-model is characterized simply by
the choice of a particular $\Omega$ projector. It is important to notice that the action is only gauge invariant with 
respect to the bosonic gauge group $G$ with Lie algebra $\mathfrak{f}^{(0)}=\mathfrak{u}(1)\times \mathfrak{u}(1)$, 
hence only the components $A_{\pm}^{(0)}$ are genuine gauge fields, the other components $A_{\pm}^{(i)},\ i=1,2,3$
play the role of auxiliary spectators fields. However, for simplicity we will refer to the whole $A_{\pm}$ as the gauge field.

In the sigma model limit, which is defined by expanding the group-like Lagrange multiplier near 
the identity $\mathcal{F=}1+\kappa ^{2}\nu /k+...$
with $k\rightarrow \infty $ and $\kappa ^{2}$ fixed, we find that%
\begin{equation}
\Omega =1+\frac{\kappa ^{2}}{k}\theta +...,\text{ \ \ }\theta =P^{(2)}+\frac{%
3}{2}P^{(1)}+\frac{1}{2}P^{(3)}. \label{omega small}
\end{equation}%
In this limit, the deformed action reduces to the action of the hybrid
superstring written in the first order (or non-Abelian Buscher) form%
\begin{equation}
S_{hybrid}=-\frac{\kappa ^{2}}{\pi }\dint_{\Sigma }d^{2}x\left\langle
A_{+}\theta A_{-}+\nu F_{+-}\right\rangle +...,  \label{hybrid action}
\end{equation}%
where the ellipsis denote sub-leading terms of order $1/k.$ After using the
equations of motion for the Lagrange multiplier field $\nu $ and by fixing the
gauge $A_{\pm }=J_{\pm }=f^{-1}\partial _{\pm }f,$ we recover the usual $%
AdS_{2}\times S^{2}$ hybrid superstring action functional\footnote{Notice the presence of a kinetic term along fermionic coset directions which otherwise is absent in the GS formalism.} \cite{Hybrid} 
\begin{equation}
S_{hybrid}=-\frac{\kappa ^{2}}{\pi }\dint_{\Sigma }d^{2}x\big\langle
(J_{+}-J_{+}^{(0)})(J_{-}-J_{-}^{(0)})-\frac{1}{2}%
(J_{+}^{(1)}J_{-}^{(3)}-J_{+}^{(3)}J_{-}^{(1)})\big\rangle .
\label{Hybrid}
\end{equation}
Alternatively, the gauge field equations of motion are given by%
\begin{equation}
A_{+}=\left( \Omega ^{T}-D^{T}\right) ^{-1}\mathcal{F}^{-1}\partial _{+}%
\mathcal{F},\text{ \ \ }A_{-}=-\left( \Omega -D\right) ^{-1}\partial _{-}%
\mathcal{FF}^{-1},\text{ \ \ }D=Ad_{\mathcal{F}}.  \label{gauge field eom}
\end{equation}%
After putting them back into the action (\ref{k-deformed action}), a
deformation of the Non Abelian T-dual of the hybrid action (\ref{Hybrid}) is
produced $S_{eff}=S'_{hybrid}+S_{WZ}+S_{dil}$, with
\begin{equation}
S'_{hybrid}=-\frac{k}{2\pi }(\lambda ^{-4}-1)\int\nolimits_{\Sigma
}d^{2}x\big\langle (\hat{J}_{+}-\hat{J}_{+}^{(0)})(\hat{J}%
_{-}-\hat{J}_{-}^{(0)})+(\hat{J}_{+}\partial _{-}\mathcal{FF}^{-1}-\partial
_{+}\mathcal{FF}^{-1}\hat{J}_{-})\big\rangle \label{hybrid'}
\end{equation}%
and where we have introduced the hatted currents%
\begin{equation}
\widehat{J}_{\pm }=(\Omega ^{T}-D^{T})^{-1}\mathcal{F}^{-1}\partial _{\pm }%
\mathcal{F}.
\end{equation}
A dilaton field is also generated in this process because the action
functional is quadratic in the fields $A_{\mu }.$ However, its explicit form
is not required for the present level of analysis but its general $\Omega $%
-dependent form can be found in \cite{k-def fer}. The combination of the $B$-fields coming from
the integration and the WZ term are such that the equations of motion are preserved.
In \cite{lambda back,lambda ads3xs3}, an
explicit construction of the background fields in the Green-Schwarz
formulation for the supercosets $AdS_{2}\times S^{2}$ and $AdS_{3}\times S^{3}$ 
is presented, there the dilaton receives contributions
from the fermionic directions of the auxiliary fields after integration.

By defining the deformed dual currents%
\EQ{
I_{\pm }^{(0)} &=A_{\pm }^{(0)},\text{ \ \ }I_{+}^{(1)}=\lambda
^{-1/2}A_{+}^{(1)},\text{ \ \ }I_{-}^{(1)}=\lambda ^{-3/2}A_{-}^{(1)}, \\
I_{\pm }^{(2)} &=\lambda ^{-1}A_{\pm }^{(2)},\text{ \ \ }%
I_{+}^{(3)}=\lambda ^{-3/2}A_{+}^{(3)},\text{ \ \ }I_{-}^{(3)}=\lambda
^{-1/2}A_{-}^{(3)}, 
\label{Dual currents} 
}%
where the $A_{\pm }$ are given by (\ref{gauge field eom}), the equations of
motion of the action (\ref{k-deformed action}), for generic values of $%
\lambda ,$ becomes exactly those of the hybrid superstring 
\EQ{
\partial _{+}I_{-}^{(0)}-\partial
_{-}I_{+}^{(0)}+[I_{+}^{(0)},I_{-}^{(0)}]+[I_{+}^{(1)},I_{-}^{(3)}]+[I_{+}^{(2)},I_{-}^{(2)}]+[I_{+}^{(3)},I_{-}^{(1)}] &=0,
 \\
D_{+}^{(0)}I_{-}^{(3)}+[I_{+}^{(1)},I_{-}^{(2)}]+[I_{+}^{(2)},I_{-}^{(1)}]
&=0,   \\
D_{-}^{(0)}I_{+}^{(1)}-[I_{+}^{(2)},I_{-}^{(3)}]-[I_{+}^{(3)},I_{-}^{(2)}]
&=0,   \\
D_{+}^{(0)}I_{-}^{(2)}+[I_{+}^{(1)},I_{-}^{(1)}] &=0,\text{ } \\
\text{\ }D_{-}^{(0)}I_{+}^{(2)}-[I_{+}^{(3)},I_{-}^{(3)}] &=0,   \\
D_{+}^{(0)}I_{-}^{(1)} &=0,\text{ \ }   \\
D_{-}^{(0)}I_{+}^{(3)} &=0, 
\label{k-def eom} 
}
where $D_{\pm }^{(0)}(\ast )=\partial _{\pm }(\ast )+[I_{\pm }^{(0)},\ast ]$
is a covariant derivative. The last two equations states that $I_{-}^{(1)}$
and $I_{+}^{(3)}$ are covariantly chiral. The whole set of equations (\ref%
{k-def eom}) can be condensed into a Lax pair representation given by 
\begin{equation}
\mathscr{L}%
_{+}(z)=I_{+}^{(0)}+zI_{+}^{(1)}+z^{2}I_{+}^{(2)}+z^{3}I_{+}^{(3)},\text{ \
\ }\mathscr{L}%
_{-}(z)=I_{-}^{(0)}+z^{-3}I_{-}^{(1)}+z^{-2}I_{-}^{(2)}+z^{-1}I_{-}^{(3)},
\label{Light-cone Lax}
\end{equation}%
which is valued in the twisted Lie superalgebra (\ref{loop superalgebra}). Under the action of the $%
%TCIMACRO{\U{2124} }%
%BeginExpansion
\mathbb{Z}
%EndExpansion
_{4}$ grading automorphism (\ref{auto}), the Lax pair satisfy%
\begin{equation}
\Phi (\mathscr{L}_{\pm }(z))=\mathscr{L}_{\pm }(iz).  \label{automorphism}
\end{equation}

There are also two currents defined by%
\begin{equation}
\mathscr{J}_{+}=-\frac{k}{2\pi }\left( \mathcal{F}^{-1}\partial _{+}\mathcal{%
F+F}^{-1}A_{+}\mathcal{F-}A_{-}\right) ,\text{ \ \ }\mathscr{J}_{-}=\frac{k}{%
2\pi }\left( \partial _{-}\mathcal{F\mathcal{F}}^{-1}\mathcal{-F}A_{-}%
\mathcal{F}^{-1}\mathcal{+}A_{+}\right)  \label{KM currents off}
\end{equation}%
that obey the algebra of two mutually commuting Kac-Moody superalgebras\footnote{%
We use $\eta _{AB}=\left\langle T_{A},T_{B}\right\rangle $, $C_{12}=\eta
^{AB}T_{A}\otimes T_{B}$ and denote $\delta_{xy}=\delta(x-y)$, $\delta^{\prime }_{xy}=
\partial _{x}\delta(x-y)$. See also the appendix \ref{A} for the tensor index convention. }  
\begin{equation}
\{\overset{1}{\mathscr{J}_{\pm }}(x),\overset{2}{\mathscr{J}_{\pm }}(y)\}=%
\frac{1}{2}[C_{12},\overset{1}{\mathscr{J}_{\pm }}(x)-\overset{2}{\mathscr{J}%
_{\pm }}(y)]\delta _{xy}\mp \frac{k}{2\pi }C_{12}\delta _{xy}^{\prime },%
\text{ \ \ }\{\overset{1}{\mathscr{J}_{\pm }}(x),\overset{2}{\mathscr{J}%
_{\mp }}(y)\}=0  \label{super KM algebra}
\end{equation}%
no matter what $\Omega $ is. They are universal to all $\lambda $-models.
On-shell, in the sense that the equations (\ref{gauge field eom}) are used, they reduce to%
\begin{equation}
\mathscr{J}_{+}=-\frac{k}{2\pi }\left( \Omega ^{T}A_{+}-A_{-}\right) ,\text{
\ \ }\mathscr{J}_{-}=-\frac{k}{2\pi }\left( \Omega A_{-}-A_{+}\right) .
\label{On-shell KM}
\end{equation}%
By defining the special values $z_{\pm }=\lambda ^{\mp 1/2}$ of the spectral
parameter$,$ we find that 
\begin{equation}
\mathscr{L}_{+}(z_{+})=\Omega ^{T}A_{+},\ \ \mathscr{L}_{-}(z_{+})=A_{-},\ \ 
\mathscr{L}_{+}(z_{-})=A_{+},\ \ \mathscr{L}_{-}(z_{-})=\Omega A_{-}.
\end{equation}%
Then, the spatial component of the Lax pair, which is defined by $\mathscr{L}%
_{1}=\mathscr{L}_{+}-\mathscr{L}_{-},\ $imply the interesting relation between the Lax operator and the Kac-Moody currents 
\begin{equation}
\mathscr{L}_{1}(z_{+})=-\frac{2\pi }{k}\mathscr{J}_{+},\text{ \ \ }\mathscr{L%
}_{1}(z_{-})=\frac{2\pi }{k}\mathscr{J}_{-}.  \label{Lax at poles}
\end{equation}%
From this we see that (set $\mathscr{L}_{1}=\mathscr{L}$ to avoid clutter)
the Kac-Moody algebra\footnote{%
After imposing the gauge field equations of motion, the Kac-Moody algebra
for the currents (\ref{On-shell KM}) is the same of (\ref{super KM algebra}%
). This is a consequence of the protection mechanism \cite{k-def bos}.} can
be written as%
\begin{equation}
\{\overset{1}{\mathscr{L}}(z_{\pm }),\overset{2}{\mathscr{L}}(z_{\pm })\}=-[%
\mathfrak{s}_{12}(z_{\pm }),\overset{1}{\mathscr{L}}(z_{\pm })-\overset{2}{%
\mathscr{L}}(z_{\pm })]\delta _{xy}-2\mathfrak{s}_{12}(z_{\pm })\delta
_{xy}^{\prime },\text{ \ \ }\mathfrak{s}_{12}(z_{\pm })=\pm \frac{\pi }{k}%
C_{12}.  \label{Maillet at poles}
\end{equation}%
One of the goals below is to find operators $\mathfrak{r}/\mathfrak{s}$ such
that the Maillet bracket is obeyed%
\begin{equation}
\{\overset{1}{\mathscr{L}}(x,z),\overset{2}{\mathscr{L}}(y,w)\}=[\mathfrak{r}%
_{12},\overset{1}{\mathscr{L}}(x,z)+\overset{2}{\mathscr{L}}(y,w)]\delta
_{xy}-[\mathfrak{s}_{12},\overset{1}{\mathscr{L}}(x,z)-\overset{2}{\mathscr{L%
}}(y,w)]\delta _{xy}-2\mathfrak{s}_{12}\delta _{xy}^{\prime },
\label{Maillet}
\end{equation}%
and that reduces to (\ref{Maillet at poles}) at the special points $z_{\pm }$.%

After integrating out the gauge fields, the effective $\lambda $-model
action is invariant under the parity-like transformation defined by%
\begin{equation}
\Pi (\mathcal{F},\Omega ,k)=(\mathcal{F}^{-1},\Omega ^{-1},-k), \label{parity-like}
\end{equation}%
whose effect on the on-shell Kac-Moody currents is to swap them 
\begin{equation}
\Pi \mathscr{J}_{\pm }=\mathscr{J}_{\mp }.
\end{equation}%
This last result follows from the identities%
\begin{equation}
\Pi A_{+}=\Omega ^{T}A_{+}.\text{ \ \ }\Pi A_{-}=\Omega A_{-}.
\end{equation}
The action of $\Pi$ as given in \eqref{parity-like} is an important symmetry common to all known $\lambda$-models (just change the $\Omega$ right above in each case). For instance, it can be exploited to constraint the very form of the $\lambda$-beta functions \cite{beta1,beta2}.

The action also has a couple of global Poisson-Lie symmetries with conserved
charges \cite{quantum group}%
\begin{equation}
m(z_{\pm })=P\exp [\pm \frac{2\pi }{k}\int\nolimits_{S^{1}}dx\mathscr{J}%
_{\pm }(x)]
\label{LR LP}
\end{equation}%
associated to the global right/left actions $\delta_{R}\mathcal{F}=\mathcal{F}\epsilon_{R}$ and
$\delta_{L}\mathcal{F}=\epsilon_{L}\mathcal{F}$ of the group $F$, respectively. These charges
are alternatively extracted by evaluating the monodromy matrix %
\begin{equation}
m(z)=P\exp [-\int\nolimits_{S^{1}}dx\mathscr{L(}x,z\mathcal{)]}
\label{left mono}
\end{equation}%
at the special points $z=z_{\pm}$ and using \eqref{Lax at poles}.
However, these two symmetries are not independent because
they obey the relation $\Pi m(z_{\pm })=m(z_{\mp })$ and play a very important role
at the quantum level when putting them in the lattice \cite{quantum group} indicating the presence of
a quantum group symmetry.

\subsection{From Noether to Poisson-Lie symmetry}\label{N and LP}

One of the main properties of the deformation is that it promotes the
global Noether symmetry of the parent $\sigma$-model associated to the left action of the group $F$
to a global Poisson-Lie group symmetry in the $\lambda$-model\footnote{For a detailed study of this type of symmetry in the case of the $\eta$-models, see \cite{Poisson-Lie eta}. } \cite{part II}. The actions \eqref{Hybrid} and \eqref{k-deformed action} have the same Lax pair structure but for the former in \eqref{Light-cone Lax} we replace $I_{\pm}$ by the left-invariant currents $J_{\pm}=f^{-1}\partial_{\pm}f$. Then, both theories have the same associated linear system\footnote{For simplicity we drop the $\overline{x}=(t,x)$ dependence on the quantities, if needed.}, namely
\EQ{(\partial_{\pm}+\mathscr{L_{\pm}(}z))\Psi(z)=0,
}
where $\Psi(z)$ is the so-called wave function.
This last equation in combination with \eqref{gauge field eom} imply that, on-shell, we have the relations\footnote{This relation is used in \cite{part I}
to construct the deformed giant magnon solutions of the $AdS_{5}\times S^{5}$ GS superstring in the lambda background.}
\EQ{f=\Psi(1)^{-1},\quad\mathcal{F}=\Psi(\lambda^{1/2})\Psi(\lambda^{-1/2})^{-1}.
\label{fields and psi}
}
Then, the constant right action of the group $F$ on the wave-function $\Psi(z)$ can be lifted to the
left action of $F$ on $f$ that leads to the well-known Noether symmetry of the $\sigma$-model generated by the charge
$Q_{L}$, i.e. we have that the variation $\delta_{X} f=Xf$, $X\in \mathfrak{f}$ can be written in the usual Abelian moment form
\EQ{\delta_{X} f(\overline{x})=\big\langle {X,\{Q_{L},f(\overline{x})\}}\big\rangle.
}
However, this action is hidden in the dual field $\mathcal{F}$ as can be seen from \eqref{fields and psi}
but it can be shown \cite{part II} that the infinitesimal right action 
$\delta\Psi(\lambda^{\pm1/2})=\Psi(\lambda^{\pm1/2})X$, $X\in \mathfrak{f}$ directly on the wave-function
can be written instead in the non-Abelian moment form
\EQ{\delta_{X}\Psi_{(\pm)}(\overline{x})=\pm\big\langle\overset{2}{X},\mathcal{\overset{\text{2}}{W}}{}^{-1}
\{\overset{2}{\mathcal{W}},\overset{1}{\Psi}_{(\pm)}(\overline{x})\}\big\rangle_{2},
}
where we have denoted $\Psi(\lambda^{\pm 1/2})=\Psi_{(\pm)}$. This shows that the global left action of $F$ in the
$\sigma$-model becomes a Poisson-Lie symmetry in the $\lambda$-model generated by the non-Abelian Hamiltonian
$\mathcal{W}$, which turns out to be the (right) monodromy matrix. It is important to stress that this kind of symmetry
only holds on-shell and that it cannot be lifted off-shell to be a symmetry of the action functional in the usual Noether theorem sense.
It is also important to notice that this situation is exactly the same for the hybrid and the Green-Schwarz formalisms and apply
as well for the charges \eqref{LR LP} extracted from the (left) monodromy matrix \eqref{left mono}. Hence, the $\lambda$-deformation naturally introduces a
$q$-deformation on the hybrid formalism.
\section{Dirac's procedure: the constraints}\label{s3}

In order to construct (\ref{Maillet}) we first need to identify which
constraints are first class and which constraints are second class, then we
need to use Dirac's procedure. However, we will only focus in identifying them
not paying attention to the specific values of the Lagrange multipliers fixed by
stability of the constraints and so on.

For the purpose of applying the Dirac procedure, we will need the Kac-Moody
current algebra written above in (\ref{super KM algebra}) plus the basic
Poisson brackets

\begin{equation}
\left\{ P_{\pm }^{A}(x),A_{\mp }^{B}(y)\right\} =\frac{1}{2}\eta ^{AB}\delta
_{xy},\ \ \partial _{0}\varphi (x)=\int dy\left\{ \varphi (x), H(y)\right\} ,
\end{equation}
where $P_{\mp }$ is the momentum field conjugate to the gauge field $A_{\pm}$.

\textbf{Step} \textbf{I. }\textit{Find the primary constraints and construct
the total Hamiltonian}. The primary constraints are given by%
\begin{equation}
P_{+}\approx 0,\text{ \ \ }P_{-}\approx 0
\end{equation}%
and the total Hamiltonian density is%
\begin{equation}
H_{T}=H_{C}-2\left\langle u_{+}P_{-}+u_{-}P_{+}\right\rangle ,
\end{equation}%
where $u_{\pm }$ are arbitrary Lagrange multipliers and%
\begin{equation}
H_{C}=-\frac{k}{\pi }\big\langle \left( \frac{\pi }{k}\right) ^{2}\left( 
\mathscr{J}_{+}^{2}+\mathscr{J}_{-}^{2}\right) -\frac{2\pi }{k}\left( A_{+}%
\mathscr{J}_{-}+A_{-}\mathscr{J}_{+}\right) +\frac{1}{2}\left(
A_{+}^{2}+A_{-}^{2}\right) -A_{+}\Omega A_{-}\big\rangle
\end{equation}%
is the canonical Hamiltonian.

\textbf{Step II. }\textit{Find the secondary constraints using }$H_{T}$%
\textit{, i.e, all relations that are }$u_{\pm }$\textit{-independent}.
There are only two secondary constraints, they are%
\begin{equation}
C_{+}=\mathscr{J}_{+}+\frac{k}{2\pi }\left( \Omega ^{T}A_{+}-A_{-}\right)
\approx 0,\text{ \ \ }C_{-}=\mathscr{J}_{-}+\frac{k}{2\pi }\left( \Omega
A_{-}-A_{+}\right) \approx 0  \label{secondary const}
\end{equation}%
and are completely equivalent to the $A_{\pm }$ equations of motion
written above in (\ref{gauge field eom}).

The symmetric stress tensor of the action (\ref{k-deformed action}) is found
(after re-installation of the world-sheet metric) by the variation of the
action with respect to the 2d metric. It has the following non-zero
components%
\begin{equation}
T_{\pm \pm }=-\frac{k}{4\pi }\big\langle (\mathcal{F}^{-1}D_{\pm }\mathcal{F%
})^{2}+2A_{\pm }(\Omega -1)A_{\pm }\big\rangle ,
\end{equation}%
where $D_{\pm }(\ast )=\partial _{\pm }(\ast )+[A_{\pm },\ast ]$ is a covariant derivative.
After using the definitions (\ref{KM currents off}), we can show that (set $%
C_{0}=C_{+}+C_{-},$ $C_{1}=C_{+}-C_{-}$)%
\begin{equation}
T_{++}+T_{--}=H_{C}-\left\langle A_{0}C_{0}\right\rangle ,
\end{equation}%
where%
\begin{equation}
T_{\pm \pm }=-\frac{k}{\pi }\big\langle \left( \frac{\pi }{k}\right) ^{2}
\mathscr{J}_{\pm }^{2}\pm \frac{\pi }{k}\mathscr{J}_{\pm }A_{1}+\frac{1}{4}%
A_{1}^{2}+\frac{1}{2}A_{\pm }(\Omega -1)A_{\pm }\big\rangle .
\end{equation}%
Alternatively, we get the relation
\begin{equation}
H_{C}=T_{++}+T_{--}+\left\langle A_{0}C_{0}\right\rangle ,
\label{canonical Hamiltonian}
\end{equation}%
where%
\EQ{
T_{++} &=-\big\langle \frac{\pi }{k}C_{+}^{2}-C_{+}(\Omega ^{T}-1)A_{+}+%
\frac{k}{4\pi }A_{+}(\Omega \Omega ^{T}-1)A_{+}\big\rangle , \\
T_{--} &=-\big\langle \frac{\pi }{k}C_{-}^{2}-C_{-}(\Omega -1)A_{-}+\frac{k%
}{4\pi }A_{-}(\Omega ^{T}\Omega -1)A_{-}\big\rangle ,  
\label{stress tensor}
}
which expresses the canonical Hamiltonian $H_{C}$ in terms of constraints
only. Of course, on the constraint surface we have that%
\begin{equation}
T_{\pm \pm }\approx -\frac{k}{4\pi }(\lambda ^{-4}-1)\big\langle A_{\pm
}^{(2)}A_{\pm }^{(2)}+2A_{\pm }^{(1)}A_{\pm }^{(3)}\big\rangle ,
\label{stress-tensor}
\end{equation}%
where the $A_{\pm }^{\prime }s$ are now determined by the conditions $C_{\pm
}\approx 0$ in terms of the other fields (\ref{gauge field eom}). The
Virasoro constraints (more specifically $T_{\pm \pm }\approx 0$) are
secondary constraints which appear as the stability conditions to the
primary constraints given by the momentum conjugates to the 2d world-sheet
metric components\footnote{In the hybrid superstring the Virasoro constraints 
are not imposed in the same way as they are imposed on the GS approach. Instead, they are
implemented through a BRST operator.}. Notice that using the equations of motion (\ref{k-def eom}%
) we can confirm that the stress-tensor components are indeed chiral for any value of $\lambda$
\begin{equation}
\partial _{\mp }T_{\pm \pm }=0\rightarrow T_{\pm \pm }=T_{\pm \pm }(x^{\pm
}).
\end{equation}

By introducing the extended Hamiltonian 
\begin{equation}
H_{E}=H_{C}-2\left\langle u_{+}P_{-}+u_{-}P_{+}+\mu _{+}C_{-}+\mu
_{-}C_{+}\right\rangle
\end{equation}%
and by forcing the stability conditions $\partial _{0}C_{\pm }\approx 0,$ we
determine all the Lagrange multipliers $u_{\pm }^{\prime }s$ but $u_{\pm
}^{(0)}$, which are linked to the bosonic gauge symmetry present in the
hybrid theory and generated by the constraint $C_{0}^{(0)}$ belonging to the
grade zero part $\mathfrak{f}^{(0)}$ of the superalgebra $\mathfrak{f}$. 
Now we have identified the full set of constraints and multipliers and hence
the algorithm stops. The last step deals with the information we have found.

\textbf{Step III.} \textit{Separate the first and second class constraints.}
There is only one primary first class constraint and it is given by $%
P_{0}^{(0)}.$ To find the secondary first class ones we must find a way to get
rid of the gauge field. The obvious combinations with the less number of
gauge field components are%
\EQ{
C &=C_{+}+\Omega ^{T}C_{-}=\mathscr{J}_{+}+\Omega^{T}\mathscr{J}_{-}+\frac{%
k}{2\pi }\left( \Omega ^{T}\Omega -1\right) A_{-},   \\
\overline{C} &=\Omega C_{+}+C_{-}=\Omega \mathscr{J}_{+}+\mathscr{J}_{-}+%
\frac{k}{2\pi }\left( \Omega \Omega ^{T}-1\right) A_{+}.  
\label{constraints}
}
From this we realize that along the supercoset directions $\mathfrak{f}%
^{(1)},\mathfrak{f}^{(2)}$ and $\mathfrak{f}^{(3)}$ all the constraints are
second class mimicking the ordinary sigma model on bosonic cosets \cite{k-def bos}. This is to be contrasted 
with the GS formulation in which along the fermionic directions $\mathfrak{f}^{(1)}$ and $\mathfrak{f%
}^{(3)}$ there is a mixture of first class and second class constraints, the
first class constraint being associated to the kappa symmetry \cite{part II}.

Then, we have found the following constraint splitting:

\underline{\textit{First class constraints}}\textit{:}%
\begin{equation}
P_{0}^{(0)},\ \ C_{0}^{(0)}=\mathscr{J}_{+}^{(0)}+\mathscr{J}_{-}^{(0)}.
\label{first class}
\end{equation}

\underline{\textit{Second class constraints}}\textit{:}%
\begin{equation}
\begin{aligned}
P_{1}^{(0)}&\text{ \ \ and \ \ }C_{-}^{(0)}, \\
(P_{+}^{(1)},P_{+}^{(2)},P_{+}^{(3)})&\text{ \ \ and \ \ }%
(C^{(1)},C^{(2)},C^{(3)}), \\
(P_{-}^{(1)},P_{-}^{(2)},P_{-}^{(3)})&\text{ \ \ and \ \ }(\overline{C}%
^{(1)},\overline{C}^{(2)},\overline{C}^{(3)}).
\end{aligned}
\end{equation}

By virtue of the protection mechanism \cite{k-def bos}, we can set all
second class constraints strongly to zero and continue using the super
Kac-Moody algebra (\ref{super KM algebra}) at no harm. Then, we are now able
to use (in the strong sense) the phase space relations%
\EQ{
A_{1}^{(0)} &=\frac{2\pi }{k}\mathscr{J}_{-}^{(0)},   \\
A_{\pm }^{(1)} &=\alpha (\lambda ^{\mp 1}\mathscr{J}_{\pm }^{(1)}+\lambda
^{2}\mathscr{J}_{\mp }^{(1)}),   \\
A_{\pm }^{(2)} &=\alpha (\mathscr{J}_{\pm }^{(2)}+\lambda ^{2}\mathscr{J}%
_{\mp }^{(2)}),   \\
A_{\pm }^{(3)} &=\alpha (\lambda ^{\pm 1}\mathscr{J}_{\pm }^{(3)}+\lambda
^{2}\mathscr{J}_{\mp }^{(3)}),  
\label{currents}
}
where we have defined%
\begin{equation}
\alpha =-\frac{2\pi }{k}\frac{1}{(z_{+}^{4}-z_{-}^{4})}.
\end{equation}%
The Poisson algebra generated by the currents (\ref{Dual currents}), i.e. 
the current algebra, is found by means of (\ref{currents}) and the Kac-Moody 
algebra structure of the theory. Their algebra is written at extend in the appendix \ref{A}.

\section{Integrability: the Maillet $\mathfrak{r/s}$ bracket}\label{s4}

In order to construct the Maillet bracket (\ref{Maillet}), we first impose strongly
all the second class constraints (\ref{currents}) on the spatial component
of the Lax connection $\mathscr{L}_{1}$ defined by (\ref{Light-cone Lax})
and second we extend it outside the constraint surface by adding to it the
only first class constraint left behind (\ref{first class}). Then, the Hamiltonian
or extended Lax operator%
\begin{equation}
\mathscr{L}^{\prime }(z)=\mathscr{L}(z)+\rho (z)C_{0}^{(0)},
\end{equation}%
is entirely expressed in terms of the components of the Kac-Moody currents $\mathscr{J%
}_{\pm }$. The function $\rho (z)$ is completely arbitrary but it must be
such that $\mathscr{L}^{\prime }(z)$ obeys (\ref{automorphism}). However, it
can be fixed by requiring that the relations
\begin{equation}
\mathscr{L}^{\prime }(z_{+})=-\frac{2\pi }{k}\mathscr{J}_{+},\text{ \ \ }%
\mathscr{L}^{\prime }(z_{-})=\frac{2\pi }{k}\mathscr{J}_{-}
\end{equation}%
are still valid outside the constraint surface. This last requirement imply that%
\begin{equation}
\rho (z)=\alpha (z^{4}-z_{-}^{4}).
\end{equation}%
Hence, we find the extended or Hamiltonian Lax operator%
\begin{equation}
\begin{aligned}
\mathscr{L}^{\prime }(z)=\alpha (z^{4}-z_{-}^{4})&\left\{ \mathscr{J}%
_{+}^{(0)}+\frac{z_{+}^{3}}{z^{3}}\mathscr{J}_{+}^{(1)}+\frac{z_{+}^{2}}{%
z^{2}}\mathscr{J}_{+}^{(2)}+\frac{z_{+}}{z}\mathscr{J}_{+}^{(3)}\right\}\\
&+\alpha (z^{4}-z_{+}^{4})\left\{ \mathscr{J}_{-}^{(0)}+\frac{z_{-}^{3}}{z^{3}%
}\mathscr{J}_{-}^{(1)}+\frac{z_{-}^{2}}{z^{2}}\mathscr{J}_{-}^{(2)}+\frac{%
z_{-}}{z}\mathscr{J}_{-}^{(3)}\right\} .  \label{extended Lax hybrid}
\end{aligned}
\end{equation}%
Of course, by construction it satisfies the property%
\begin{equation}
\Phi (\mathscr{L}^{\prime }(z))=\mathscr{L}^{\prime }(iz).
\end{equation}

Notice that if we extend the action of the omega projector (\ref{projector}) $\Omega $ to the
whole complex plane by defining%
\begin{equation}
\Omega (z)=P^{(0)}+z^{-3}P^{(1)}+z^{-2}P^{(2)}+z^{-1}P^{(3)},
\end{equation}%
where obviously $\Omega =\Omega (\lambda ),$ and use the identities $\Omega
(z)\Omega (w)=\Omega (w)\Omega (z)=\Omega (zw)$, we can write quite compactly%
\begin{equation}
\mathscr{L}^{\prime }(z)=f_{-}(z)\Omega \left( z/z_{+}\right) \mathscr{J}%
_{+}+f_{+}(z)\Omega \left( z/z_{-}\right) \mathscr{J}_{-},\text{ \ \ }f_{\pm
}(z)=\alpha (z^{4}-z_{\pm }^{4}).  \label{Hybrid Lax projector}
\end{equation}%
We can profit from this notation and write the Maillet bracket in terms of $%
(z,\lambda )$-dependent projectors acting on the Kac-Moody superalgebras%
\begin{eqnarray}
\begin{aligned}
\{\overset{1}{\mathscr{L}^{\prime }}(x,z),\overset{2}{\mathscr{L}^{\prime }}%
(y,w)\}&=\!f_{-}(z)f_{-}(w)\overset{1}{\Omega }\left( z/z_{+}\right) \overset%
{2}{\Omega }\left( w/z_{+}\right) \{\overset{1}{\mathscr{J}_{+}}(x),\overset{%
2}{\mathscr{J}_{+}}(y)\}   \\
&+f_{+}(z)f_{+}(w)\overset{1}{\Omega }\left( z/z_{-}\right) \overset{2}{%
\Omega }\left( w/z_{-}\right) \{\overset{1}{\mathscr{J}_{-}}(x),\overset{2}{%
\mathscr{J}_{-}}(y)\},  
\end{aligned}
\label{Maillet as KM}
\end{eqnarray}%
which clearly satisfy the condition (\ref{Maillet at poles}) at the special
points $z_{\pm }$. We can also write (\ref{Light-cone Lax}) in the compact
form%
\begin{equation}
\mathscr{L}_{+}(z)=\Omega ^{T}\left( z_{-}/z\right) A_{+},\text{ \ \ }%
\mathscr{L}_{-}(z)=\Omega \left( z/z_{+}\right) A_{-}.
\end{equation}%
To recover the Lax operator $\mathscr{L}_{1}$ on the constraint surface with
the pair $\mathscr{L}_{\pm }$ given by (\ref{Light-cone Lax}), we simply
replace in (\ref{Hybrid Lax projector}) the on-shell values of the Kac-Moody
currents $\mathscr{J}_{\pm }$ as given in (\ref{On-shell KM}).

From the central terms of the Kac-Moody algebras we can immediately isolate
the symmetric part of the AKS $R$-matrix, namely,%
\begin{equation}
\mathfrak{s}_{12}(z,w)=\frac{k}{4\pi }\Big[ f_{-}(z)f_{-}(w)\overset{1}{%
\Omega }\left( z/z_{+}\right) \overset{2}{\Omega }\left( w/z_{+}\right)
-f_{+}(z)f_{+}(w)\overset{1}{\Omega }\left( z/z_{-}\right) \overset{2}{%
\Omega }\left( w/z_{-}\right) \Big] C_{12}.
\end{equation}%
There is a special value of the deformation parameter where $\mathfrak{s}_{12}$ simplifies%
\begin{equation}
\underset{\lambda \rightarrow 0}{\lim }\ \mathfrak{s}_{12}(z,w)=-\frac{\pi }{k}%
C_{12}^{(00)}.
\end{equation}%
This means that in this limit the non-ultralocality of the theory is
contained (or tamed or alleviated) within the grade zero part of the
superalgebra and that it is not affected by the coset directions. This same result
also holds for the GS formalism\footnote{See also \cite{vicedo} for the first attempt to alleviate the non-ultralocality of the GS superstring.} \cite{part II} and the purely bosonic theories \cite{quantum group} and 
is a version of the Faddeev-Reshetikhin ultralocalization mechanism \cite{Ultra} but now applied 
to this particular model. 

An explicit calculation reveals that%
\begin{equation}
\mathfrak{s}%
_{12}(z,w)=s_{0}C_{12}^{(00)}+s_{1}C_{12}^{(13)}+s_{2}C_{12}^{(22)}+s_{3}C_{12}^{(31)},
\end{equation}%
where%
\EQ{
s_{0}(z,w) &=-\frac{\alpha }{2}\left[ z^{4}+w^{4}-(z_{+}^{4}+z_{-}^{4})%
\right] , \\
s_{1}(z,w) &=\frac{\alpha }{2}\frac{1}{z^{3}w}\left[ 1-z^{4}w^{4}\right] ,
\\
s_{2}(z,w) &=\frac{\alpha }{2}\frac{1}{z^{2}w^{2}}\left[ 1-z^{4}w^{4}\right]
, \\
s_{3}(z,w) &=\frac{\alpha }{2}\frac{1}{zw^{3}}\left[ 1-z^{4}w^{4}\right] .
}
We can write this compactly as%
\begin{equation}
\mathfrak{s}_{12}(z,w)=-\frac{1}{z^{4}-w^{4}}\tsum\nolimits_{j=0}^{3}%
\{z^{j}w^{4-j}C_{12}^{(j,4-j)}\varphi _{\lambda
}^{-1}(w)-z^{4-j}w^{j}C_{12}^{(4-j,j)}\varphi _{\lambda }^{-1}(z)\},
\end{equation}%
where%
\begin{equation}
\varphi _{\lambda }^{-1}(z)=-2\alpha \left[ \varphi _{\sigma
}^{-1}(z)+\epsilon ^{2}(\lambda )\right] ,
\end{equation}%
is the deformed twisting function and%
\begin{equation}
\varphi _{\sigma }^{-1}(z)=\frac{1}{4}(z^{2}-z^{-2})^{2},\text{ \ \ }%
\epsilon ^{2}(\lambda )=-\frac{1}{4}(z_{+}^{2}-z_{-}^{2})^{2}.
\end{equation}
The first term above is the well-known $\sigma$-model twisting function, the second term
implements the deformation and is responsible for displacing the poles of 
$\varphi _{\sigma }(z)$ along the real axis. Note that the special values $z_{\pm}$ introduced above 
are two of the displaced poles of the original sigma model twisting function.
 
Now, using the symmetric part $\mathfrak{s}_{12}$ as an input in the Maillet bracket, we can
solve for the antisymmetric part of the $R$-matrix 
\begin{equation}
\mathfrak{r}_{12}(z,w)=\frac{1}{z^{4}-w^{4}}\tsum\nolimits_{j=0}^{3}%
\{z^{j}w^{4-j}C_{12}^{(j,4-j)}\varphi _{\lambda
}^{-1}(w)+z^{4-j}w^{j}C_{12}^{(4-j,j)}\varphi _{\lambda }^{-1}(z)\},
\end{equation}%
showing that the deformed hybrid formulation of the superstring is still integrable as it can be put in
Maillet's form. The use of the projectors in showing this is quite powerful.
All these results fit perfectly within the analysis presented in \cite{part II}.

\subsection{Relation to the $\protect\lambda $-model of the GS superstring}

To show the equivalence of the deformed hybrid (H) and Green-Schwarz (GS)
superstrings at the level of the Maillet brackets, we need to show that for
the GS case the extended Lax operator takes exactly the same form as in the
hybrid formulation (\ref{Hybrid Lax projector}) with the same projector
operator $\Omega $!. This could come as a surprise but we will show this is
indeed the case. A similar situation was realized in \cite{magro} for the
un-deformed traditional sigma models.

For the sake of comparison, we write the projectors associated to both
formulations 
\begin{eqnarray}
\Omega _{H}(z) &=&P^{(0)}+z^{-3}P^{(1)}+z^{-2}P^{(2)}+z^{-1}P^{(3)}, \\
\Omega _{GS}(z) &=&P^{(0)}+z^{-1}P^{(1)}+z^{-2}P^{(2)}+zP^{(3)}.
\end{eqnarray}%
To show the equivalence we will work in reverse instead. Suppose that (\ref%
{extended Lax hybrid}) or (\ref{Hybrid Lax projector}) is given and that we
consider the special form for the super Kac-Moody currents 
\begin{eqnarray}
\begin{aligned}
\mathscr{J}_{\pm }^{\prime } =\ &\mp \frac{k}{4\pi }\left[ (1-z_{\pm
}^{4})\Pi ^{(0)}+2\mathcal{A}^{(0)}\right] \pm \frac{k}{4\pi }\frac{z_{\pm }%
}{2}\left[ 2(1-z_{\mp }^{4})\Pi ^{(1)}-(3+z_{\mp }^{4})\mathcal{A}^{(1)}%
\right] \\
&+\frac{k}{4\pi }\left[ (z_{+}^{2}-z_{-}^{2})\Pi ^{(2)}\mp
(z_{+}^{2}+z_{-}^{2})\mathcal{A}^{(2)}\right] \mp \frac{k}{4\pi }\frac{%
z_{\mp }}{2}\left[ 2(1-z_{\pm }^{4})\Pi ^{(3)}+(3+z_{\pm }^{4})\mathcal{A}%
^{(2)}\right]
\end{aligned}
\end{eqnarray}%
in term of a new set of conjugate fields $(\mathcal{A},\Pi ).$ Then, we get that%
\begin{eqnarray}
\begin{aligned}
\mathscr{L}_{GS}^{\prime }(z) =\ &\mathcal{A}^{(0)}+\frac{1}{4}(3z+z^{-3})%
\mathcal{A}^{(1)}+\frac{1}{2}(z^{2}+z^{-2})\mathcal{A}^{(2)}+\frac{1}{4}%
(3z^{-1}+z^{3})\mathcal{A}^{(3)} \\
&+\frac{1}{2}(1-z^{4})\Pi ^{(0)}+\frac{1}{2}(z^{-3}-z)\Pi ^{(1)}+\frac{1}{2}%
(z^{-2}-z^{2})\Pi ^{(2)}+\frac{1}{2}(z^{-1}-z^{3})\Pi ^{(3)},
\end{aligned}
\end{eqnarray}%
which is nothing but the Hamiltonian GS Lax operator that takes into account the extension by the fermionic
constraints proper of the GS formalism \cite{magro,vicedo}. This
shows that both formulations has the same extended Lax operator and the same
set of $\mathfrak{r/s}$ matrices. This is because in the GS case the
arbitrary functions multiplying the fermionic constraints are arbitrary and
can be fixed by demanding equivalence to the hybrid formalism \cite{magro}.
This means that as phase spaces the hybrid formulation and the Green-Schwarz
formulation of the $AdS_{2}\times S^{2}$ superstring are the same. The only
difference being their dynamics and the local symmetries (defined through the $%
\Omega $-dependence of the constraints in (\ref{constraints})) involved. Indeed, notice
that the particular combination of projectors%
\begin{equation}
(\Omega\Omega^{T})_{H}-1=(\lambda ^{-4}-1)(P^{(1)}+P^{(2)}+P^{(3)}),%
\text{ \ \ }(\Omega\Omega^{T}) _{GS}-1=(\lambda ^{-4}-1)P^{(2)},
\end{equation}%
change dramatically the Dirac analysis of the phase space constraints. In the former case it is 
conjectured \cite{Hybrid} the existence of a quantum $N=(2,2)$ world-sheet superconformal symmetry
that replaces the kappa symmetry\footnote{ At classical level,
after gauge fixing the kappa symmetry there is a global fermionic symmetry
leftover, which in the $\lambda \rightarrow 0$ can be identified with an exotic global 2d
$(N,N)$ extended supersymmetry. The $N$ being the rank of the kappa symmetry
that was gauge fixed \cite{01,02,susy flows, inte vs susy,SSSSG}.}
(2+2 to be exact) present in the latter case, both gauge symmetries being used to remove
un-physical degrees of freedom from the spectrum. However, it is important to realize that not even the classical 
part of this superconformal symmetry (corresponding to a W-algebra) is manifest in the action fuctional, 
as can be seen from the Dirac analysis. Below we will show how to construct explicitly the classical generator
for this hidden symmetry.

\section{Deformed Poisson brackets and the $\protect\lambda \rightarrow 0$
limit}\label{s5}

The fact that the Poisson brackets of the Lax operator can be put in the
Maillet $\mathfrak{r/s}$ form, means that we can write in a compact way the
Poisson bracket for functions on $\mathcal{L}$ in terms of the usual $R$
bracket associated to the twisted loop Lie superalgebra $\widehat{\mathfrak{f%
}}.$ Namely,%
\begin{equation}
\{F,G\}(\mathscr{L}^{\prime })=(\mathscr{L}^{\prime },[dF,dG]_{R})_{\varphi
_{\lambda }}+\omega (R(dF),dG)_{\varphi _{\lambda }}+\omega
(dF,R(dG))_{\varphi _{\lambda }},  \label{R-bracket}
\end{equation}%
where $R=\pm (\Pi _{\geq 0}-\Pi _{<0})$ is the usual AKS $R$-matrix,%
\begin{eqnarray}
(X,Y)_{\varphi _{\lambda }} &=&\int_{S^{1}}dx\oint_{0}\frac{dz}{2\pi iz}%
\varphi _{\lambda }(z)\left\langle X(x,z),Y(x,z)\right\rangle , \\
\omega (X,Y)_{_{\varphi _{\lambda }}} &=&\int_{S^{1}}dx\oint_{0}\frac{dz}{%
2\pi iz}\varphi _{\lambda }(z)\left\langle X(x,z),\partial
_{1}Y(x,z)\right\rangle 
\end{eqnarray}%
are the twisted inner product and co-cycle, respectively, and\footnote{%
As a curiosity, note that $\mathscr{L}_{1}^{\prime }(z)=R\mathscr{L}%
_{0}^{\prime }(z).$}%
\begin{equation}
\mathscr{L}_{1}^{\prime
}(z)=I_{1}^{(0)}+zI_{+}^{(1)}+z^{2}I_{+}^{(2)}+z^{3}I_{+}^{(3)}-z^{-3}I_{-}^{(1)}-z^{-2}I_{-}^{(2)}-z^{-1}I_{-}^{(3)}+\rho (z)C_{0}^{(0)}
\end{equation}%
is the extended Lax operator written this time in terms of the dual currents.
Above, $\Pi _{\geq 0}$ and $\Pi _{<0}$ are projectors along positive/negative
powers of $z$ acting on quantities valued in the loop superalgebra $\widehat{\mathfrak{f}}$.

The functions on $\mathscr{L}^{\prime }$ and their associated differentials
are defined by the usual relations%
\begin{equation}
F(\mathscr{L}^{\prime })=(F,\mathscr{L}^{\prime })_{\varphi _{\lambda }},%
\text{ \ \ }\underset{t\rightarrow 0}{\lim }\ \frac{d}{dt}F(\mathscr{L}%
^{\prime }\mathcal{+}tX)=(dF,X)_{\varphi_{\lambda} }.
\end{equation}

For the current components $I_{\pm }$, we use $F(\mathscr{L}^{\prime }) =(F,\mathscr{L}^{\prime })_{\varphi _{\lambda }}$ with
\begin{eqnarray}
\begin{aligned}
F(x,z) =&\ \varphi _{\lambda }^{-1}(z)[(1+z_{-}^{4}z^{-4})\mu ^{(0)}+z^{-1}\mu
^{(3)}+z^{-2}\mu ^{(2)}+z^{-3}\mu ^{(1)}]  \\
&-\varphi _{\lambda }^{-1}(z)[z^{3}\nu ^{(3)}+z^{2}\nu ^{(2)}+z\nu ^{(1)}] \
\end{aligned}
\end{eqnarray}%
and similarly for the constraint, we use $F(\mathscr{L}^{\prime }) =(F,\mathscr{L}^{\prime })_{\varphi _{\lambda }}$ with
\begin{eqnarray}
F(x,z) &=&\frac{1}{\alpha }\varphi _{\lambda }^{-1}(z)z^{-4}\eta ^{(0)}.
\end{eqnarray}%
Above, $\mu ,\nu ,\eta :S^{1}\rightarrow $ $\widehat{\mathfrak{f}}$ are test
functions that we remove at the end of calculations. By obvious linearity, it follows that the differentials are
simply found by setting $F\rightarrow dF.$ For the positive part, i.e,
positive powers of $z$, we have%
\begin{eqnarray}
\begin{aligned}
\int_{S^{1}}dx\big\langle \eta ^{(0)},C_{0}^{(0)}\big\rangle &\rightarrow\,
\frac{1}{\alpha }\varphi _{\lambda }^{-1}(z)z^{-4}\eta ^{(0)}, \\
\int_{S^{1}}dx\big\langle \mu ^{(0)},I_{1}^{(0)}\big\rangle &\rightarrow\,
\varphi _{\lambda }^{-1}(z)(1+z_{-}^{4}z^{-4})\mu ^{(0)}, \\
\int_{S^{1}}dx\big\langle \mu ^{(3)},I_{+}^{(1)}\big\rangle &\rightarrow\,
\varphi _{\lambda }^{-1}(z)z^{-1}\mu ^{(3)}, \\
\int_{S^{1}}dx\big\langle \mu ^{(2)},I_{+}^{(2)}\big\rangle &\rightarrow\,
\varphi _{\lambda }^{-1}(z)z^{-2}\mu ^{(2)}, \\
\int_{S^{1}}dx\big\langle \mu ^{(1)},I_{+}^{(3)}\big\rangle &\rightarrow\,
\varphi _{\lambda }^{-1}(z)z^{-3}\mu ^{(1)}.
\end{aligned}
\end{eqnarray}
For the negative part, i.e, negative powers of $z$, we get%
\begin{eqnarray}
\begin{aligned}
\int_{S^{1}}dx\big\langle \nu ^{(3)},I_{-}^{(1)}\big\rangle &\rightarrow\,
-\varphi _{\lambda }^{-1}(z)z^{3}\nu ^{(3)}, \\
\int_{S^{1}}dx\big\langle \nu ^{(2)},I_{-}^{(2)}\big\rangle &\rightarrow\,
-\varphi _{\lambda }^{-1}(z)z^{2}\nu ^{(2)}, \\
\int_{S^{1}}dx\big\langle \nu ^{(1)},I_{-}^{(3)}\big\rangle &\rightarrow\,
-\varphi _{\lambda }^{-1}(z)z\nu ^{(1)}.
\end{aligned}
\end{eqnarray}%
Now it is a turn to compute the deformed current algebra for the $\lambda $%
-model of the hybrid superstring. It is written in appendix \ref{A} below, after using
the $\mathfrak{r}/\mathfrak{s}$ approach we find perfect agreement with the
more direct and pedestrian computation that follows from the relations that are consequence
of (\ref{currents}) and the Kac-moody algebra structure of the theory.

In the $\lambda \rightarrow 0$ limit a dramatic simplification of the
current algebra occurs. The only non-zero brackets being (those involving
the constraint remain the same)%
\begin{eqnarray}
\begin{aligned}
\{\overset{1}{I_{1}^{(0)}}(x),\overset{2}{I_{1}^{(0)}}(y)\} =&-\frac{2\pi }{%
k}([C_{12}^{(00)},\overset{2}{I_{1}^{(0)}}(y)]\delta
_{xy}-C_{12}^{(00)}\delta _{xy}^{\prime }), \\
\{\overset{1}{I_{1}^{(0)}}(x),\overset{2}{I_{-}^{(i)}}(y)\} =&-\frac{2\pi }{%
k}[C_{12}^{(00)},\overset{2}{I_{-}^{(i)}}(y)]\delta _{xy},\text{ \ \ }%
i=1,2,3
\end{aligned}
\end{eqnarray}%
for the brackets involving the grade zero current and %
\begin{eqnarray}
\begin{aligned}
\{\overset{1}{I_{+}^{(1)}}(x),\overset{2}{I_{+}^{(1)}}(y)\} &=\frac{2\pi }{k%
}[C_{12}^{(13)},\overset{2}{I_{+}^{(2)}}(y)]\delta _{xy}, \\
\{\overset{1}{I_{+}^{(1)}}(x),\overset{2}{I_{+}^{(2)}}(y)\} &=\frac{2\pi }{k%
}[C_{12}^{(13)},\overset{2}{I_{+}^{(3)}}(y)]\delta _{xy}, \\
\{\overset{1}{I_{-}^{(2)}}(x),\overset{2}{I_{-}^{(3)}}(y)\} &=\frac{2\pi }{k%
}[C_{12}^{(22)},\overset{2}{I_{-}^{(1)}}(y)]\delta _{xy}, \\
\{\overset{1}{I_{-}^{(3)}}(x),\overset{2}{I_{-}^{(3)}}(y)\} &=\frac{2\pi }{k%
}[C_{12}^{(31)},\overset{2}{I_{-}^{(2)}}(y)]\delta _{xy}
\end{aligned}
\end{eqnarray}%
for the currents along the coset directions. Notice that, very remarkably, the $\pm $ 
light-cone sectors along the coset directions completely decouple in the sense that the
current components $I_{\pm}$ do not mix, manifesting 2d relativistic invariance in this limit.
The theory has the same mild non-ultralocality as in the Green-Schwarz\ case 
but this time there is no Poisson Casimir and the usual
connection to the Pohlmeyer reduction, their associated generalized
sine-Gordon models and its mKdV-type integrable hierarchy typical of the GS superstring 
\cite{susy flows,SSSSG,inte vs susy,NA kinks,part II,Grigo,Andrei,revisited} 
is absent for this case, showing that the $\lambda$-deformation 
is along a different direction in the space of Poisson structures.
It would be very interesting to further explore the hybrid
formalism in the $\lambda \rightarrow 0$ limit, in particular it seems to be it might have
simpler OPE's and vertex operators as they depend on the
symplectic structure of the theory, which drastically simplifies in this
limit. Indeed, the fact that in this limit the $\pm $ sectors decouple (at least classically) suggest that (anti)-chiral objects do not mix either
raising the interesting possibility of computing exact OPE's even in a curved background.

\section{The $N=(2,2)$ superconformal algebra}\label{N=2}

The un-deformed theory (\ref{Hybrid}) is conjectured to posses an $N=(2,2)$ superconformal symmetry
at the quantum level \citep{Hybrid}. We will restrict here to a purely classical analysis and argue, however,
that it is reasonable to expect that this conjecture might be extended to the $\lambda$-model as well and this is suggested by the independence of the 
symmetry algebra structure on the deformation parameter $\lambda$. Recall that Poisson brackets only capture the information
of the OPE's that corresponds to the classical results (like single contractions), then we will only be able to reproduce the W-algebra structure of 
the theory. For further details on the conjecture see \cite{Hybrid}.

Start writing the stress tensor components (\ref{stress tensor}) in the form 
\begin{equation}
T_{\pm \pm }=\frac{1}{2\alpha}\left\langle K_{\pm },K_{\pm
}\right\rangle ,\text{ \ \ }K_{\pm }=I_{\pm }^{(2)}+I_{\pm }^{(1)}+I_{\pm
}^{(3)}.
\end{equation}%
Using the Poisson algebra written down in the appendix \ref{A}, we find that on
the constraint surface (when $C_{0}^{(0)}\approx 0$) and for any $\lambda$, we get the usual
Virasoro algebra
\begin{equation}
\{T_{\pm \pm }(x),T_{\pm \pm }(y)\}=\pm (T_{\pm \pm }^{\prime }(x)\delta
_{xy}+2T_{\pm \pm }(x)\delta _{xy}^{\prime }),
%\text{ \ \ }\{T_{\pm \pm
%}(x),T_{\mp \mp }(y)\}=0.
\end{equation}
with other brackets being zero.

To construct the classical chiral generators a more refined analysis of the Lie superalgebra
$\mathfrak{f}$ is required, see appendix \ref{B} for details. Indeed, the current components, of say 
$I_{+}^{(3)}$, decompose under the action of the gauge algebra $\mathfrak{f}^{(0)}=\mathfrak{u}(1)\times \mathfrak{u}(1)$ as follows
\EQ{I_{+}^{(3)}\rightarrow \{I_{(++)}^{(3)},I_{(--)}^{(3)},I_{(+-)}^{(3)},I_{(-+)}^{(3)}\},  
}
where we have dropped the light-cone index $+$ in favor of the gauge labels.
This decomposition imply that the two gauge invariant fermion bilinears defined by
\EQ{G^{+}=c I_{(++)}^{(3)}\cdot I_{(--)}^{(3)},\quad G^{-}=c I_{(+-)}^{(3)}\cdot I_{(-+)}^{(3)},
}
with $c$ arbitrary, satisfy the chirality condition
\EQ{
\partial_{-}G^{\pm}=0\rightarrow G^{\pm}=G^{\pm}(x^{+})
}
by virtue of the last equation of motion in \eqref{k-def eom} as the current $I_{+}^{(3)}$ is covariantly chiral. Recall that the equations of motion are the same for any value of $\lambda$.
A similar results is valid for $I_{-}^{(1)}$.

Now we proceed to compute the classical symmetry algebra for the chiral sector $(+)$. Appendices \ref{A} and \ref{B} imply
that on the constraint surface (when $C_{0}^{(0)}\approx 0$) we have the following Poisson brackets
\begin{eqnarray}
\{T_{++}(x),G^{\pm }(y)\} &=&G^{\prime \pm }(x)\delta _{xy}+2G^{\pm
}(x)\delta _{xy}^{\prime },  \notag \\
\{G^{+}(x),G^{-}(y)\}
&=&W(x)\delta_{xy},  \label{partial} \\
\{G^{\pm }(x),G^{\pm }(y)\} &=&0,  \notag
\end{eqnarray}
where we have set $c^{2}\alpha l=1$ and introduced a new generator
\EQ{W=I_{(++)}^{(3)}\cdot [I_{+}^{(2)},I_{+}^{(3)}]_{(--)}^{(1)}
+I_{(--)}^{(3)}\cdot [I_{+}^{(2)},I_{+}^{(3)}]_{(++)}^{(1)}.
}
This last generator is a spin-3 current
\EQ{
\begin{aligned}
\{T_{++}(x),W(y)\} &= 2 W^{\prime }(x)\delta
_{xy}+3W(x)\delta _{xy}^{\prime },
\end{aligned}
}
and reveals a classical W-algebra structure\footnote{This last Poisson bracket matches (up to a sign) the one introduced in \cite{W-algebra} to compute the conformal weight $\Delta$ of the $W$-current. Namely,
$\{T(x),W(y)\}=-(\Delta-1)W^{\prime }(x)\delta
_{xy}-\Delta W(x)\delta _{xy}^{\prime }$. }.
Other Poisson brackets mixing elements of different sectors vanish identically. The important point is that
this symmetry algebra is independent of the deformation parameter $\lambda$, then it is natural
to conjecture that the $\lambda$-model has the same $N=(2,2)$ superconformal symmetry 
of the original hybrid action \eqref{Hybrid}. This is because using either the dual currents $I_{\pm}$ or the original currents $J_{\pm}$,
the chiral symmetry algebras are quite the same in form and content.
\section{Conformal invariance: one-loop beta function}\label{s7}

We quickly review the calculation of the one-loop beta function of \cite%
{Hybrid} (see also \cite{Ido,zarembo} for the GS formalism) but combined with the simpler constant background current used in 
\cite{k-def beta}, in which bosonic and fermionic fluctuations decouple
making the calculation simpler. Then, we apply the same strategy to the deformed 
theory.

\subsection{The sigma model}

Consider the un-deformed hybrid action\footnote{%
For\ the choice $s=1/2,$ we recover the hybrid superstring action (\ref{Hybrid}).} given by \cite{Hybrid} 
\begin{equation}
S_{hybrid}=-\frac{\kappa ^{2}}{2\pi t_{N}}\dint_{\Sigma }d^{2}x\left\langle
J_{+},\theta J_{-}\right\rangle _{N},\text{ \ \ }\theta
=P^{(2)}+(1+s)P^{(1)}+(1-s)P^{(3)},  \label{general hybrid}
\end{equation}%
where $t_{N}$ is the Dynkin index of the defining $N$-dimensional
representation of the superalgebra $\mathfrak{f}$. See Appendix A of \cite%
{k-def beta} for further details on the Lie algebraic conventions used through this section.

The fluctuations fields to be used $\eta \in \mathfrak{f}$ are Lie superalgebra
valued and are related to the fluctuations of the currents $J_{\pm },$
through the basic relations%
\begin{equation}
\delta J_{\pm }=\frac{1}{\kappa }D_{\pm }\eta ,\text{ \ \ }f^{-1}\delta
f\equiv \frac{1}{\kappa }\eta ,\text{ \ \ and \ \ }(\delta D_{\pm })\eta
=[D_{\pm }\eta ,\eta ],
\end{equation}%
where $D_{\pm }=\partial _{\pm }+[J_{\pm },\ast ]$ is a covariant derivative$%
.$ By fixing the gauge $\eta ^{(0)}=0,$ associated to the gauge symmetry of the action,
and by using the following specific choice of background field given by%
\begin{equation}
f=\exp x^{\mu }\Theta _{\mu }\rightarrow J_{\pm }=\Theta _{\pm },
\end{equation}%
where $\Theta _{\mu }\in \mathfrak{f}^{(2)}$ are constant fields satisfying $%
[\Theta _{\mu },\Theta _{\nu }]=0,$ we find the operators that govern the
fluctuations $\eta $. Namely, 
\begin{equation}
\mathcal{D}_{B}(x)=\left( \mathcal{-\partial }_{+}\partial _{-}+\Theta
_{+}\Theta _{-}\right) \text{ \ \ acting on \ \ }\eta ^{(2)}
\end{equation}%
for the bosonic sector and%
\begin{equation}
\mathcal{D}_{F}(x)=\left( 
\begin{array}{cc}
\mathcal{-\partial }_{+}\partial _{-}+s\Theta _{+}\Theta _{-} & (s-\frac{1}{2%
})\Theta _{+}\partial _{-}-(s+\frac{1}{2})\Theta _{-}\partial _{+} \\ 
\mathcal{(}s-\frac{1}{2}\mathcal{)}\Theta _{-}\partial _{+}-\mathcal{(}s+%
\frac{1}{2}\mathcal{)}\Theta _{+}\partial _{-} & \mathcal{-\partial }%
_{+}\partial _{-}-s\Theta _{+}\Theta _{-}%
\end{array}%
\right) \text{\ acting on \ }\left( 
\begin{array}{c}
\text{\ }\eta ^{(1)} \\ 
\text{\ }\eta ^{(3)}%
\end{array}%
\right)
\end{equation}%
for the fermionic sector. Notice that this is basically the content of the eq. (4.16) of \cite%
{Hybrid} after some obvious re-arrangements and identifications.

After Wick rotating and gathering all logarithmic divergences, we find the
one-loop contribution to the effective Lagrangian in Euclidean signature 
\cite{Hybrid}
\begin{equation}
I_{undef}^{1-loop}=-\frac{1}{8\pi }\ln \mu \cdot \lbrack
Tr_{adj}^{(0)}+Tr_{adj}^{(2)}-(2s^{2}+\frac{1}{2}%
)(Tr_{adj}^{(1)}+Tr_{adj}^{(3)})](\Theta \cdot \Theta ).
\end{equation}%
The theory (\ref{general hybrid}) has vanishing one-loop beta function \cite%
{Hybrid} precisely when $s=\pm 1/2$ as a consequence of the vanishing of the
dual Coxeter number or, equivalently, the quadratic Casimir operator in the adjoint representation
of $\mathfrak{f}=\mathfrak{psu}(1,1|2)$.

\subsection{The lambda model}

Now we want to discover if the deformation described by (\ref{k-deformed action}) preserves the
one loop conformal invariance of the action \eqref{Hybrid}. Recall that our choice above 
corresponds to $s=1/2.$ 

After using the gauge field $A_{\pm }$ equations of motion, we obtain the effective lambda model action\footnote{%
The dilaton contribution $S_{dil}$ is not necessary at this level of
analysis as we are interested only in the quantum scale invariance, not Weyl.} (cf. \eqref{hybrid'})
\begin{equation}
S_{\lambda }=-\frac{k}{4\pi t_{N}}\dint_{\Sigma }d^{2}x\left\langle \mathcal{%
F}^{-1}\partial _{+}\mathcal{F[}1+2(\Omega -D)^{-1}D\mathcal{]F}%
^{-1}\partial _{-}\mathcal{F}\right\rangle _{N}+S_{WZ}+S_{dil}.
\label{effective action}
\end{equation}%
Using the background field (compared with $f$ above)%
\begin{equation}
\mathcal{F}=\exp x^{\mu }\Lambda _{\mu },
\end{equation}%
where $\Lambda _{\mu }\in \mathfrak{f}^{(2)}$ are constant fields satisfying 
$[\Lambda _{\mu },\Lambda _{\nu }]=0,$ we get the following dual background
currents%
\begin{equation}
I_{\pm }^{(2)}\equiv \Theta _{\pm }=\pm \frac{\lambda }{(1-\lambda ^{2})}%
\Lambda _{\pm },\text{ \ \ }I_{\pm }^{(i)}=0,\text{ \ \ }i=0,1,3.
\label{deformed back field}
\end{equation}%
From the equations of motion (\ref{k-def eom}), we obtain the operators
governing the fluctuations of the bosonic and fermionic sectors. For the
bosonic sector we get%
\begin{equation}
\mathcal{D}_{B}(x)=\left( 
\begin{array}{cccc}
\partial _{-} & 0 & 0 & -\Theta _{+} \\ 
0 & \partial _{+} & -\Theta _{-} & 0 \\ 
-\Theta _{-} & \Theta _{+} & -\partial _{-} & \partial _{+} \\ 
0 & 0 & \partial _{-} & \partial _{+}%
\end{array}%
\right) \text{ \ \ acting on \ \ }\left( 
\begin{array}{c}
\widehat{I}_{+}^{(2)} \\ 
\widehat{I}_{-}^{(2)} \\ 
\widehat{I}_{+}^{(0)} \\ 
\widehat{I}_{-}^{(0)}%
\end{array}%
\right) ,
\end{equation}%
where the last line right above is the analogue of the gauge fixing
condition $\eta ^{(0)}=0$ used in the un-deformed hybrid sigma model. For the
fermionic sector, we obtain

\begin{equation}
\mathcal{D}_{F}(x)=\left( 
\begin{array}{cccc}
\partial _{-} & 0 & \Theta _{-} & -\Theta _{+} \\ 
0 & \partial _{+} & 0 & 0 \\ 
0 & 0 & \partial _{-} & 0 \\ 
-\Theta _{-} & \Theta _{+} & 0 & \partial _{+}%
\end{array}%
\right) \text{ \ \ acting on \ \ }\left( 
\begin{array}{c}
\widehat{I}_{+}^{(1)} \\ 
\widehat{I}_{-}^{(1)} \\ 
\widehat{I}_{+}^{(3)} \\ 
\widehat{I}_{-}^{(3)}%
\end{array}%
\right) .
\end{equation}

The 1-loop quantum effective Lagrangian in Euclidean signature is then given
by%
\begin{equation}
\mathcal{L}_{E}^{eff}=\mathcal{L}_{E}^{(0)}+I_{def}^{1-loop},\text{ \ \ }%
I_{def}^{1-loop}=\frac{1}{2}\int\nolimits_{|p|<\mu }\frac{d^{2}p}{(2\pi )^{2}%
}tr[\log \mathcal{D}_{B}(p)-\log \mathcal{D}_{F}(p)],
\end{equation}%
where%
\begin{equation}
\mathcal{L}_{E}^{(0)}=\frac{k}{16\pi t_{N}}\frac{1+\lambda ^{2}}{1-\lambda
^{2}}\left\langle \Lambda \cdot \Lambda \right\rangle _{N}
\end{equation}%
and%
\begin{equation}
\mathcal{D}_{B}(p)=\left( 
\begin{array}{cccc}
p_{-} & 0 & 0 & -\Theta _{+} \\ 
0 & p_{+} & -\Theta _{-} & 0 \\ 
-\Theta _{-} & \Theta _{+} & -p_{-} & p_{+} \\ 
0 & 0 & p_{-} & p_{+}%
\end{array}%
\right) ,\text{ \ \ }\mathcal{D}_{F}(p)=\left( 
\begin{array}{cccc}
p_{-} & 0 & \Theta _{-} & -\Theta _{+} \\ 
0 & p_{+} & 0 & 0 \\ 
0 & 0 & p_{-} & 0 \\ 
-\Theta _{-} & \Theta _{+} & 0 & p_{+}%
\end{array}%
\right) .
\end{equation}%
The contributions associated to logarithmic divergences (denoted by the symbol $\doteq $%
) are%
\begin{eqnarray}
\frac{1}{2}\int\nolimits_{|p|<\mu }\frac{d^{2}p}{(2\pi )^{2}}tr[\log 
\mathcal{D}_{B}(p)] &\doteq &-\frac{1}{8\pi }\ln \mu \lbrack
Tr_{adj}^{(0)}+Tr_{adj}^{(2)}](\Theta \cdot \Theta ), \\
-\frac{1}{2}\int\nolimits_{|p|<\mu }\frac{d^{2}p}{(2\pi )^{2}}tr[\log 
\mathcal{D}_{F}(p)] &\doteq &\frac{1}{8\pi }\ln \mu \lbrack
Tr_{adj}^{(1)}+Tr_{adj}^{(3)}](\Theta \cdot \Theta ).
\end{eqnarray}%
Altogether we get, to one-loop in $1/k$ but exact in $\lambda$, that%
\begin{equation}
I_{def}^{1-loop}=-\frac{1}{8\pi }\ln \mu \cdot \lbrack
Tr_{adj}^{(0)}+Tr_{adj}^{(2)}-(Tr_{adj}^{(1)}+Tr_{adj}^{(3)})](\Theta \cdot
\Theta ),
\end{equation}%
which is proportional to the un-deformed one-loop contribution found above.
The proportionality factor being $\lambda $-dependent, determined by (\ref%
{deformed back field}) and can be found by writing $\Theta _{\pm }$ in terms
of $\Lambda _{\pm }$. The $\lambda $-deformation besides preserving the
underlying integrability of the original hybrid superstring sigma model also
preserves its 1-loop conformal invariance, i.e. the coupling $\lambda$ is marginal to this order.
It would be very interesting to follow the lines of \cite{lambda back,lambda ads3xs3} and to verify if
this $\lambda $-model is also Weyl invariant at quantum level by constructing explicitly the background
fields and checking if they obey the relevant set of supergravity equations of motion.

\section{Digression on the pure spinor $\protect\lambda $-model}\label{s8}
In this last final section we speculate on the possibility of associating a $\lambda$-model to the pure 
spinor (PS) superstring on the background $AdS_{5}\times S^{5}$. One of the main properties of the PS 
formalism \cite{PS,ICTP} is that on it both the kappa symmetry and the Virasoro constraints characteristics of the GS formalism
are replaced by a single BRST symmetry. Fortunately, by demanding BRST invariance of the $\lambda$-model of the hybrid superstring
plus a term involving the PS ghosts we are able to construct a sensible deformation. Unfortunately, the deformation
does not seem to preserve integrability.

\subsection{The $\protect\sigma $-model of the PS superstring}

The pure spinor superstring action on $AdS_{5}\times S^{5}$ is given by%
\begin{equation}
S=-\frac{\kappa ^{2}}{\pi }\dint_{\Sigma } d^{2}x\left\langle J_{+}\theta
J_{-}\right\rangle -\frac{2\kappa ^{2}}{\pi }\dint_{\Sigma } d^{2}x\big\langle
w^{(3)}D_{-}^{(0)}l^{(1)}+\overline{w}^{(1)}D_{+}^{(0)}\overline{l}^{(3)}+N%
\overline{N}\big\rangle , \label{PS action}
\end{equation}%
where $D_{\pm }^{(0)}(\ast )=\partial _{\pm }(\ast )+\big[ J_{\pm
}^{(0)},\ast \big] $ is a covariant derivative,  $J_{\pm }=f^{-1}\partial _{\pm }f$ is the usual
flat current and $\theta$ is the same projector used for the hybrid superstring \eqref{omega small}.
This time the Lie superalgebra to be considered is $\mathfrak{psu}(2,2|4)$.    
Now, $l^{(1)}$ and $\overline{l}^{(3)}$ are ghosts satisfying the pure spinor constraints%
\begin{equation}
\big[ l^{(1)},l^{(1)}\big] _{+}=\big[ \overline{l}^{(3)},\overline{l}%
^{(3)}\big] _{+}=0 \label{PS constraints}
\end{equation}%
and $w^{(3)}$ and $\overline{w}^{(1)}$ are their conjugate momenta. It is
important to notice that $l^{(1)}$ and $\overline{l}^{(3)}$ are fermionic in
character because%
\begin{equation}
l^{(1)}=l^{\alpha }T_{\alpha },\text{ \ \ }\overline{l}^{(3)}=l^{\overline{%
\alpha }}T_{\overline{\alpha }}\text{,}
\end{equation}%
where $T_{\alpha }\in \mathfrak{f}^{(1)},$ $T_{\overline{\alpha }}\in 
\mathfrak{f}^{(3)}$ are fermionic generators of $\mathfrak{f}$, while the components $%
l^{\alpha }$ and $l^{\overline{\alpha }}$ are bosonic spinors. Also%
\begin{equation}
N=-\left[ l^{(1)},w^{(3)}\right] _{+},\text{ \ \ }\overline{N}=-\big[ 
\overline{l}^{(3)},\overline{w}^{(1)}\big] _{+},
\end{equation}%
are the PS Lorentz currents. They are bosonic and belong to $\mathfrak{f}^{(0)}.$

The action \eqref{PS action} is invariant under an on-shell BRST symmetry
\begin{equation}
\delta _{B}f=f(l^{(1)}+\overline{l}^{(3)}),\text{ \ \ }\delta _{B}\overline{w%
}^{(1)}=-J_{-}^{(1)},\text{ \ \ }\delta _{B}w^{(3)}=-J_{+}^{(3)},\text{ \ \ }%
\delta _{B}l^{(1)}=\delta _{B}\overline{l}^{(3)}=0 \label{BRST sigma}
\end{equation}%
and it is also classical integrable \cite{breno,magro} with a Lax pair given by
\begin{equation}
\begin{aligned}
\mathcal{L}_{+}(z)
&=J_{+}^{(0)}+zJ_{+}^{(1)}+z^{2}J_{+}^{(2)}+z^{3}J_{+}^{(3)}+(z^{4}-1)N, \\
\mathcal{L}_{-}(z)
&=J_{-}^{(0)}+z^{-3}J_{-}^{(1)}+z^{-2}J_{-}^{(2)}+z^{-1}J_{-}^{(3)}+(z^{-4}-1)%
\overline{N}.
\end{aligned}
\end{equation}
Now, we proceed to deform this theory.

\subsection{The ``$\protect\lambda $-model'' of the PS superstring}

In order to construct the lambda model of the pure spinor superstring, we
need to find a way to: I) preserve its BRST symmetry and II) preserve
its integrability. Our strategy will be to start with I) and later verify if II) is guaranteed by the resulting deformation.

We construct the lambda model for the PS
superstring by adding to the $\lambda $-deformed hybrid action
\begin{equation}
S_{hybrid}=S_{F/F}(\mathcal{F},A_{\mu })-\frac{k}{\pi }%
\dint_{\Sigma }d^{2}x\left\langle A_{+}(\Omega -1)A_{-}\right\rangle \label{hybrid},
\end{equation}%
a term proportional to the PS ghosts

\begin{equation}
S_{ghost}=r\dint d^{2}x\big\langle w^{(3)}D_{-}^{(0)}l^{(1)}+\overline{w}%
^{(1)}D_{+}^{(0)}\overline{l}^{(3)}+sN\overline{N}\big\rangle ,
\end{equation}%
where $r,s$ are parameters to be determined by BRST symmetry arguments and $D_{\pm }^{(0)}(\ast
)=\partial _{\pm }(\ast )+\big[ A_{\pm }^{(0)},\ast \big]$ is a covariant derivative. Namely, we define
\begin{equation}
S_{PS}=S_{hybrid}+S_{ghost}. \label{PS}
\end{equation}

In order to find a candidate BRST symmetry we will work on stages. Start by considering the matter part and propose
(set $\delta _{B}=\overline{\delta }$) the following transformations\footnote{This is the same method used in
\cite{k-def fer} to construct the kappa symmetry for the GS case. Notice the resemblance between the kappa and the BRST symmetry in both formulations.}
\begin{equation}
\overline{\delta }\mathcal{F}=-\alpha \mathcal{F}+\mathcal{F} \beta ,\text{ \
\ }\overline{\delta }A_{+}=D_{+}\alpha ,\text{ \ \ \ }\overline{\delta }%
A_{-}=D_{-}\beta, 
\end{equation}%
where $\alpha $ and $\beta $ are functions of $l^{(1)},$ $\overline{l}^{(3)}.
$ The variation of the first term in \eqref{hybrid} is given by%
\begin{equation}
\overline{\delta }S_{F/F}=\frac{k}{\pi }\dint_{\Sigma }d^{2}x\big\langle
\left( \alpha -\beta \right) F_{+-}\big\rangle , \label{F/F}
\end{equation}%
where $F_{+-}\neq 0$ is the curvature of $A_{\pm }$. The variation of the $\Omega $%
-dependent part of \eqref{hybrid} is of the form%
\begin{eqnarray}
\begin{aligned}
\overline{\delta }S_{\Omega }
=&\ \frac{k}{\pi}(\lambda -1)\dint_{\Sigma }d^{2}x\big\langle 
-cl^{(1)}F_{+-}^{(3)}+b\overline{l}^{(3)}F_{+-}^{(1)}\big\rangle 
\\
&+ \frac{k}{\pi }\lambda (\lambda
^{-4}-1)\dint_{\Sigma }d^{2}x\big\langle cl^{(1)}D_{-}^{(0)}A_{+}^{(3)}+b\overline{l}%
^{(3)}D_{+}^{(0)}A_{-}^{(1)}\big\rangle ,
\end{aligned}
\end{eqnarray}
where we have taken%
\begin{equation}
\alpha =\lambda cl^{(1)}+b\overline{l}^{(3)},\text{ \ \ }\beta
=cl^{(1)}+\lambda b\overline{l}^{(3)},
\end{equation}%
with $b$ and $c$ arbitrary constants. Using this particular choice we end up
with%
\begin{equation}
\overline{\delta }S_{hybrid}=\frac{k}{\pi }\lambda (\lambda
^{-4}-1)\dint_{\Sigma }d^{2}x\big\langle cl^{(1)}D_{-}^{(0)}A_{+}^{(3)}+b%
\overline{l}^{(3)}D_{+}^{(0)}A_{-}^{(1)}\big\rangle .
\end{equation}%
From this expression, we discover that by taking $r=-\frac{k}{\pi }(\lambda ^{-4}-1)$ and setting%
\begin{equation}
\overline{\delta }\overline{w}^{(1)}=-\lambda bA_{-}^{(1)},\text{ \ \ }%
\overline{\delta }w^{(3)}=-\lambda cA_{+}^{(3)},\text{ \ \ }\overline{\delta 
}l^{(1)}=\overline{\delta }\overline{l}^{(3)}=0,
\end{equation}%
we obtain for the whole action that%
\begin{equation}
\overline{\delta }S_{PS}=\frac{k}{\pi }(\lambda
^{-4}-1)\dint_{\Sigma }d^{2}x\big\langle \overline{N}\overline{\delta }
(A_{+}^{(0)}-sN) +N\overline{\delta }(A_{-}^{(0)}-s%
\overline{N}) \big\rangle .
\end{equation}%
By setting $s=1$, we arrive at the desired form%
\begin{equation}
\overline{\delta }S_{PS}=\frac{k}{\pi }(\lambda
^{-4}-1)\dint_{\Sigma }d^{2}x\big\langle bA_{+}^{(1)}\big[ 
\overline{l}^{(3)},\overline{N}\big]+cA_{-}^{(3)}\big[ l^{(1)},N\big]\big\rangle.
\end{equation}%
Finally, we notice that the action is BRST invariant $\overline{\delta }S_{PS}=$ $0$ because, say
\begin{equation}
\left[ l^{(1)},N\right] =\frac{1}{2}\big[ w^{(3)},\left[ l^{(1)},l^{(1)}%
\right] _{+}\big],
\end{equation}%
vanishes by virtue of the pure spinor constraints \eqref{PS constraints}.
This is where the formulation borrows its name. 

In the sigma model limit $\lambda \rightarrow 1$, the action \eqref{PS} reduces to the first order form%
\begin{equation}
S_{PS}=-\frac{\kappa ^{2}}{\pi }\dint_{\Sigma }d^{2}x\left\langle
A_{+}\theta A_{-}+\nu F_{+-}\right\rangle -\frac{2\kappa ^{2}}{\pi }%
\dint_{\Sigma }d^{2}x\big\langle w^{(3)}D_{-}^{(0)}l^{(1)}+\overline{w}%
^{(1)}D_{+}^{(0)}\overline{l}^{(3)}+N\overline{N}\big\rangle,
\end{equation}%
which is to be compared with \eqref{PS action}.
However, after taking the limit the action is no longer BRST invariant
because in this limit $\alpha = \beta$ and hence $\overline{\delta }\left\langle \nu F_{+-}\right\rangle
=0$. When compared with \eqref{F/F} this term is needed to cancel some contributions of the curvature 
coming from the variation of $S_{\Omega}$. Only when we use the $\nu $ 
equations of motion and fix the gauge $A_{\pm }=f^{-1}\partial _{\pm }f,$ the BRST symmetry is restored, i.e. when
we return to the original formulation and to the set of variations \eqref{BRST sigma}.

Now, we can find the equations of motion by varying the action we have constructed. For the field $\mathcal{F}$, we get
\begin{equation}
\begin{aligned}
\delta S_{PS} &=-\frac{k}{\pi }\dint_{\Sigma }d^{2}x\left\langle 
\mathcal{F}^{-1}\delta \mathcal{F}\left[ \partial _{+}+\Omega
^{T}A_{+}+(\lambda ^{-4}-1)N,\partial _{-}+A_{-}\right] \right\rangle  \\
&=-\frac{k}{\pi }\dint_{\Sigma }d^{2}x\left\langle \delta \mathcal{FF}^{-1}\left[
\partial _{+}+A_{+},\partial _{-}+\Omega A_{-}+(\lambda ^{-4}-1)\overline{N}%
\right] \right\rangle , \label{equivalent}
\end{aligned}
\end{equation}%
after using the $A_{\pm }$ equations of motion
\begin{equation}
\begin{aligned}
A_{+}&=\left( \Omega ^{T}-D^{T}\right) ^{-1}[\mathcal{F}^{-1}\partial _{+}%
\mathcal{F}-(\lambda^{-4}-1)N], \\
A_{-}&=-\left( \Omega -D\right) ^{-1}[\partial _{-}%
\mathcal{FF}^{-1}+(\lambda^{-4}-1)\overline{N}].  \label{A eom}
\end{aligned}
\end{equation}%
The ghosts have the same equations as in the un-deformed theory
\begin{equation}
D_{+}^{(0)}\overline{N}+[\overline{N},N]=0,\text{ \ \ }D_{-}^{(0)}N+[N,%
\overline{N}]=0.
\end{equation}

If the deformation is to preserve the integrability, then the two expressions \eqref{equivalent} for the equations of
motion should be equivalent to the evaluation of the curvature 
\begin{equation}
\left[ \partial _{+}+\mathcal{L}_{+}(z),\partial _{-}+\mathcal{L}_{-}(z)%
\right] ,
\end{equation}%
of the Lax pair
\begin{equation}
\begin{aligned}
\mathcal{L}_{+}(z)
&=I_{+}^{(0)}+zI_{+}^{(1)}+z^{2}I_{+}^{(2)}+z^{3}I_{+}^{(3)}+(z^{4}-1)N^{%
\prime }, \\
\mathcal{L}_{-}(z)
&=I_{-}^{(0)}+z^{-3}I_{-}^{(1)}+z^{-2}I_{-}^{(2)}+z^{-1}I_{-}^{(3)}+(z^{-4}-1)%
\overline{N}^{\prime }
\end{aligned}
\end{equation}%
at the special values of the spectral paramenter $z=\lambda ^{-1/2}$ and $z=\lambda ^{1/2}$, respectively\footnote{This 
time the currents $I_{\pm}$ will include contributions from the PS currents.}. The prime
in $N^{\prime }$ and $\overline{N}^{\prime }$ is to denote possible
re-scalings of the ghosts in terms of the parameter $\lambda $ similar to the ones required to define the
currents $I_{\pm }.$ We conclude that under the present (naive) construction, the pure spinor
superstring does not seem to admit a $\lambda $-model and more work is to be required. A possibility is to add a
new term in order to restore integrability. This new term should, in principle, possess
the following properties:
\begin{itemize}
\item It must be BRST invariant and gauge invariant, at least under the
gauge group generated by the $\mathfrak{f}^{(0)}$ part of the Lie
superalgebra $\mathfrak{f=psu}(2,2|4),$
\item It must be become a sub-leading correction of the order $O(1/k)$ in the sigma
model limit $\lambda \rightarrow 1$, where $k \rightarrow \infty$ with $\kappa^{2}$ fixed.
\end{itemize}
By replacing \eqref{A eom} back into \eqref{PS} we find that the resulting effective action differs from
\eqref{hybrid'} plus the ghosts term action by a non-standard coupling between the currents $\hat{J}_{\pm}$'s and $N,\overline{N}$. Another hint that perhaps we need to add a new term in order to compensate the extra terms. However, we will leave this problem to be considered more carefully in a companion paper.

\section{Concluding remarks}\label{conclusions}

In this paper we have studied in detail the $\lambda$-model of the hybrid formalism
of the superstring in the background\footnote{The supercoset $AdS_{3}\times S^{3}$ can 
be treated along the same lines.} $AdS_{2}\times S^{2}$ and showed how it preserves most
of the main characteristics of the original $\sigma$-model except the one related to the
maximal isometry group of the target space, a situation that is common to all $\lambda$-models.
The presence of Poisson-Lie groups at classical level is a strong signal
of a quantum group symmetry $F_{q}$, which should appear as the symmetry group of some non-commutative space\footnote{
This is certainly an interesting situation to be further explored in the context of the $AdS/CFT$ duality.
For an example of this in relation to the $\eta$-deformation see \cite{Tongeren}.}.
From the point of view of string theory the claim is that in the $\lambda$-model the isometry group $F$ of the
original target space is replaced by $F_{q}$ with $q$ a phase but
without breaking any of the conditions that makes the deformed target space a genuine 
string background\footnote{As initially suggested by the vanishing of the beta functions.}.
This is also supported by the recent results of \cite{lambda back,lambda ads3xs3,ads5xs5}
showing that the background fields judiciously extracted from the $\lambda$-model action functional form a one-parameter family of solutions 
to the supergravity equations of motion of relevance to each case.  
In the present situation the target space metric has fermionic directions as well and the explicit construction
of it should be much more involved than in the Green-Schwarz formalism. We leave the problem
of the explicit construction of the background fields for the near future. 

One of the goals of this work was to gain a better understanding of the structure of 
the $\lambda$-deformation itself, in the sense of clarifying its true content from the
integrable systems point of view. For further work devoted to this specific question see the papers 
\cite{part I,part II} to which the present results should be added as a complement.

Relying on our findings and on what is known for the GS superstring on 
$AdS_{5}\times S^{5},$ it is reasonable to expect that the $\lambda $-model
for the GS superstring on the supercoset $AdS_{4}\times \mathbb{C}P^{3}$
is not only classical integrable but also one-loop conformal invariant.
This can be seen from the Lie algebraic properties of semi-symmetric spaces
\cite{zarembo} and from the fact that there is no difference in the construction of the Lagrangian in comparison to the
case of $AdS_{5}\times S^{5}$. The Lax pair representation is also the same \cite{ad4xcp3} and as a consequence of this the determinant  for the fluctuations will be proportional to the quadratic Casimir in the adjoint representation as well.

Finally, an interesting question to be considered is if the Poisson-Lie T-duality that is known
between the $\eta$ and the $\lambda$ models of the Green-Schwarz superstring has an analogue for the hybrid superstring as well, i.e, if the action \eqref{Hybrid} admits a deformation of the Yang-Baxter type in terms of an $R$-matrix
satisfying the cmYBE as constructed in \cite{eta ads5xs5} for the GS formalism. However, it is already known that the $\eta$-deformation is not Weyl invariant at the quantum level for the GS case and perhaps an analogue situation could be present in the hybrid superstring as well. A possible way out of this situation in both formulations might be to consider Yang-Baxter deformations in terms of dynamical $R$-matrices instead of the usual constant ones. Hopefully, they could be general enough
as to introduce the necessary freedom required to restore Weyl's symmetry. We will come back to this question elsewhere.

\section*{Acknowledgements}

%The work of TJH is supported by...\\
%The work of JLM is supported by...\\
The work of DMS is supported by the FAPESP post-doc grant: 2012/09180-9. \\
DMS thanks Nathan Berkovits and Andrei Mikhailov for their comments and suggestions. Special thanks to T. J. Hollowood and J. L. Miramontes for valuable discussions and collaboration. The author would like to thank the referee for very useful suggestions.

\appendix
\appendixpage

These two appendices gather the most useful algebraic results used for calculations in the body of the paper:
The deformed current algebra and the $\mathfrak{psu}(1,1|2)$ Lie superalgebra proper to the $AdS_{2}\times S^{2}$ supercoset.

\section{Current algebra for the deformed hybrid formulation}\label{A}

The non-zero Poisson brackets for the currents (\ref{Dual currents}) or (\ref%
{currents}) can be computed directly from\footnote{%
The $R$ with the minus sign is the one that reproduce the Poisson brackets
computed directly from (\ref{currents}).}\eqref{R-bracket} by using the identities%
\begin{eqnarray}
\begin{aligned}
\{\left\langle \mu ,I_{\alpha }\right\rangle ,\left\langle \overline{\mu }%
,I_{\beta }\right\rangle \} =&\big\langle \{\overset{1}{I_{\alpha}}
,\overset{2}{I_{\beta}}\},(\mu \otimes \overline{\mu })\big\rangle _{12},
\\
\left\langle \lbrack \mu ,\overline{\mu }],I_{\alpha }\right\rangle
=&-\big\langle [C_{12},\overset{2}{I_{\alpha}}],\mu \otimes \overline{\mu }%
\big\rangle _{12}, \\
\left\langle \mu ,\overline{\mu }\right\rangle =&\left\langle C_{12},\mu
\otimes \overline{\mu }\right\rangle _{12},
\end{aligned}
\end{eqnarray}%
where $\mu ,\overline{\mu }$ are the test functions and $\alpha ,\beta =\pm$.
The upper indices $1,2$ refer to the copy
in a chain of tensor products. For example, $\overset{1}{u}=u\otimes I,\ 
\overset{2}{u}=I\otimes u$. The lower indices $1,2$ indicate taking the supertrace 
on the first or on the second copy of the vector space in the tensor product. 

The non-zero current algebra elements are given by 
\begin{eqnarray}
\begin{aligned}
\{\overset{1}{I_{1}^{(0)}}(x),\overset{2}{I_{1}^{(0)}}(y)\} &=-\frac{2\pi }{%
k}([C_{12}^{(00)},\overset{2}{I_{1}^{(0)}}(y)]\delta
_{xy}-C_{12}^{(00)}\delta _{xy}^{\prime }), \\
\{\overset{1}{I_{1}^{(0)}}(x),\overset{2}{I_{\pm }^{(1)}}(y)\} &=\pm \alpha
\lbrack C_{12}^{(00)},\overset{2}{I_{\mp }^{(1)}}(y)-z_{\mp }^{4}\overset{2}{%
I_{\pm }^{(1)}}(y)]\delta _{xy}, \\
\{\overset{1}{I_{1}^{(0)}}(x),\overset{2}{I_{\pm }^{(2)}}(y)\} &=\pm \alpha
\lbrack C_{12}^{(00)},\overset{2}{I_{\mp }^{(2)}}(y)-z_{\mp }^{4}\overset{2}{%
I_{\pm }^{(2)}}(y)]\delta _{xy}, \\
\{\overset{1}{I_{1}^{(0)}}(x),\overset{2}{I_{\pm }^{(3)}}(y)\} &=\pm \alpha
\lbrack C_{12}^{(00)},\overset{2}{I_{\mp }^{(3)}}(y)-z_{\mp }^{4}\overset{2}{%
I_{\pm }^{(3)}}(y)]\delta _{xy},
\end{aligned}
\end{eqnarray}%
for $[\mathfrak{f}^{(0)},\mathfrak{f}^{(i)}]$, $i=0,1,2,3$.
\begin{eqnarray}
\begin{aligned}
\{\overset{1}{I_{+}^{(1)}}(x),\overset{2}{I_{+}^{(1)}}(y)\} &=\alpha
\lbrack C_{12}^{(13)},\overset{2}{I_{-}^{(2)}}(y)-a\overset{2}{I_{+}^{(2)}}%
(y)]\delta _{xy}, \\
\{\overset{1}{I_{\pm }^{(1)}}(x),\overset{2}{I_{-}^{(1)}}(y)\} &=-\alpha
\lbrack C_{12}^{(13)},\overset{2}{I_{\pm }^{(2)}}(y)]\delta _{xy}, \\
\{\overset{1}{I_{+}^{(1)}}(x),\overset{2}{I_{+}^{(2)}}(y)\} &=\alpha
\lbrack C_{12}^{(13)},\overset{2}{I_{-}^{(3)}}(y)-a\overset{2}{I_{+}^{(3)}}%
(y)]\delta _{xy}, \\
\{\overset{1}{I_{+}^{(1)}}(x),\overset{2}{I_{-}^{(2)}}(y)\} &=-\alpha
\lbrack C_{12}^{(13)},\overset{2}{I_{+}^{(3)}}(y)]\delta _{xy}, \\
\{\overset{1}{I_{-}^{(1)}}(x),\overset{2}{I_{\pm }^{(2)}}(y)\} &=-\alpha
\lbrack C_{12}^{(13)},\overset{2}{I_{\pm }^{(3)}}(y)]\delta _{xy}, \\
\{\overset{1}{I_{\pm }^{(1)}}(x),\overset{2}{I_{\pm }^{(3)}}(y)\} &=\mp
\alpha ([C_{12}^{(13)},\overset{2}{I_{1}^{(0)}}(y)\pm \alpha z_{\pm
}^{4}C_{0}^{(0)}(y)]\delta _{xy}-C_{12}^{(13)}\delta _{xy}^{\prime }), \\
\{\overset{1}{I_{\pm }^{(1)}}(x),\overset{2}{I_{\mp }^{(3)}}(y)\} &=-\alpha
^{2}[C_{12}^{(13)},\overset{2}{C_{0}^{(0)}}(y)]\delta _{xy},
\end{aligned}
\end{eqnarray}%
for $[\mathfrak{f}^{(1)},\mathfrak{f}^{(i)}]$, $i=1,2,3$.
\begin{eqnarray}
\begin{aligned}
\{\overset{1}{I_{\pm }^{(2)}}(x),\overset{2}{I_{\pm }^{(2)}}(y)\} &=\mp
\alpha ([C_{12}^{(22)},\overset{2}{I_{1}^{(0)}}(y)\pm \alpha z_{\pm
}^{4}C_{0}^{(0)}(y)]\delta _{xy}-C_{12}^{(22)}\delta _{xy}^{\prime }), \\
\{\overset{1}{I_{+}^{(2)}}(x),\overset{2}{I_{-}^{(2)}}(y)\} &=-\alpha
^{2}[C_{12}^{(22)},\overset{2}{C_{0}^{(0)}}(y)]\delta _{xy}, \\
\{\overset{1}{I_{+}^{(2)}}(x),\overset{2}{I_{\pm }^{(3)}}(y)\} &=-\alpha
\lbrack C_{12}^{(22)},\overset{2}{I_{\pm }^{(1)}}(y)]\delta _{xy}, \\
\{\overset{1}{I_{-}^{(2)}}(x),\overset{2}{I_{+}^{(3)}}(y)\} &=-\alpha
\lbrack C_{12}^{(22)},\overset{2}{I_{-}^{(1)}}(y)]\delta _{xy}, \\
\{\overset{1}{I_{-}^{(2)}}(x),\overset{2}{I_{-}^{(3)}}(y)\} &=\alpha
\lbrack C_{12}^{(22)},\overset{2}{I_{+}^{(1)}}(y)-a\overset{2}{I_{-}^{(1)}}%
(y)]\delta _{xy},
\end{aligned}
\end{eqnarray}%
for $[\mathfrak{f}^{(2)},\mathfrak{f}^{(i)}]$, $i=2,3$ and
\begin{eqnarray}
\begin{aligned}
\{\overset{1}{I_{+}^{(3)}}(x),\overset{2}{I_{\pm }^{(3)}}(y)\} &=-\alpha
\lbrack C_{12}^{(31)},\overset{2}{I_{\pm }^{(2)}}(y)]\delta _{xy}, \\
\{\overset{1}{I_{-}^{(3)}}(x),\overset{2}{I_{-}^{(3)}}(y)\} &=\alpha
\lbrack C_{12}^{(31)},\overset{2}{I_{+}^{(2)}}(y)-a\overset{2}{I_{-}^{(2)}}%
(y)]\delta _{xy},
\end{aligned}
\end{eqnarray}%
for $[\mathfrak{f}^{(3)},\mathfrak{f}^{(3)}]$. We have defined $a\equiv z_{+}^{4}+z_{-}^{4}$. 

Notice that $a=-2(2\epsilon ^{2}-1)$ for comparison with previous works that make 
use of $\epsilon ^{2}=-\frac{(1-\lambda^{2})^{2}}{4\lambda^{2}}$ as the deformation 
parameter. In the sigma model limit when $\lambda \rightarrow 1$, the Poisson brackets 
above coincide with the current algebra of the matter sector of the pure spinor 
superstring computed in \cite{Kluson}.

Finally, the brackets involving the gauge constraint
are the standard ones 
\begin{eqnarray}
\begin{aligned}
\{\overset{1}{C_{0}^{(0)}}(x),\overset{2}{C_{0}^{(0)}}(y)\}
&=-[C_{12}^{(00)},\overset{2}{C_{0}^{(0)}}(y)]\delta _{xy}, \\
\{\overset{1}{C_{0}^{(0)}}(x),\overset{2}{I_{1}^{(0)}}(y)\}
&=-([C_{12}^{(00)},\overset{2}{I_{1}^{(0)}}(y)]\delta
_{xy}-C_{12}^{(00)}\delta _{xy}^{\prime }), \\
\{\overset{1}{C_{0}^{(0)}}(x),\overset{2}{I_{\pm }^{(i)}}(y)\}
&=-[C_{12}^{(00)},\overset{2}{I_{\pm }^{(i)}}(y)]\delta _{xy},\text{ \ \ }%
i=1,2,3.
\end{aligned}
\end{eqnarray}

\section{A basis for the $\mathfrak{psu}(1,1|2)$ Lie superalgebra}\label{B}

For completeness we write the basis presented in \cite{superparticle} and include the fermionic
elements used here to construct explicitly the $W$-element related to the superconformal 
algebra in section \ref{N=2}.

The (anti)-commutation relations are (the $m$'s are even while the $q$'s are odd)
\EQ{
%\begin{eqnarray*}
\begin{aligned}
\lbrack m_{\text{ }\beta }^{\alpha },m_{\text{ }\delta }^{\gamma }]
&=\delta _{\text{ }\beta }^{\gamma }m_{\text{ }\delta }^{\alpha }-\delta _{%
\text{ }\delta }^{\alpha }m_{\text{ }\beta }^{\gamma },\text{ \, \qquad }[%
\overline{m}_{\text{ }j}^{i},\overline{m}_{\text{ }n}^{k}]=\delta _{\text{\ }%
j}^{k}\overline{m}_{\text{ }n}^{i}-\delta _{\text{ }n}^{i}\overline{m}_{%
\text{ }j}^{k}, \\
\lbrack m_{\text{ }\beta }^{\alpha },q_{\text{ }\gamma }^{k}] &=-\delta _{%
\text{ }\gamma }^{\alpha }q_{\text{ }\beta }^{k}+\frac{1}{2}\delta _{\text{ }%
\beta }^{\alpha }q_{\text{ }\gamma }^{k},\text{ \qquad }[m_{\text{ }\beta
}^{\alpha },\overline{q}_{\text{ }k}^{\gamma }]=\delta _{\text{ }\beta
}^{\gamma }\overline{q}_{\text{ }k}^{\alpha }-\frac{1}{2}\delta _{\text{ }%
\beta }^{\alpha }\overline{q}_{\text{ }k}^{\gamma }, \\
\lbrack \overline{m}_{\text{\ }j}^{i},q_{\text{ }\alpha }^{k}] &=\delta _{%
\text{\ }j}^{k}q_{\text{ }\alpha }^{i}-\frac{1}{2}\delta _{\text{\ }j}^{i}q_{%
\text{ }\alpha }^{k},\text{ \ \ \ \qquad }[\overline{m}_{\text{ }j}^{i},\overline{q}_{%
\text{ }k}^{\alpha }]=-\delta _{\text{ }k}^{i}\overline{q}_{\text{\ }%
j}^{\alpha }+\frac{1}{2}\delta _{\text{\ }j}^{i}\overline{q}_{\text{ }%
k}^{\alpha }, \\
\lbrack q_{\text{ }\gamma }^{i},\overline{q}_{\text{\ }j}^{\beta }]_{+}
&=l(\delta _{\text{\ }j}^{i}m_{\text{\ }\alpha }^{\beta }+\delta _{\text{ }%
\beta }^{\alpha }\overline{m}_{\text{ }j}^{i}),\text{ \qquad  }l^{2}=-1,
%\end{eqnarray*}%
\end{aligned}
}
where $\alpha ,\beta =1,2$ and $i,j=1,2.$ The bosonic subalgebras $\mathfrak{su}%
(1,1)$ and $\mathfrak{su}(2)$ are generated by $m_{\text{ }\beta }^{\alpha }$
and $\overline{m}_{\text{ }j}^{i}$, respectively. There are 8 supercharges
$q_{\text{ }\gamma }^{k}$, $\overline{q}_{\text{ }%
k}^{\gamma }$.

Under the $%
%TCIMACRO{\U{2124} }%
%BeginExpansion
\mathbb{Z}
%EndExpansion
_{4}$ decomposition $\mathfrak{f}=\bigoplus\nolimits_{i=0}^{3}\mathfrak{f}%
^{(i)}$, the generators split as follows%
\EQ{
\begin{aligned}
\mathfrak{f}^{(0)} &=span\{m_{\text{ }1}^{1},\text{ }\overline{m}_{\text{ }%
1}^{1}\}, \\
\mathfrak{f}^{(1)} &=span\{q_{\text{ }1}^{1},\text{ }q_{\text{ }2}^{2},%
\text{ }\overline{q}_{\text{ }2}^{1},\text{ }\overline{q}_{\text{ }1}^{2}\},
\\
\mathfrak{f}^{(2)} &=span\{m_{\text{ }2}^{1},\text{ }m_{\text{ }1}^{2},%
\text{ }\overline{m}_{\text{ }2}^{1},\text{ }\overline{m}_{\text{ }1}^{2}\},
\\
\mathfrak{f}^{(3)} &=span\{\overline{q}_{\text{ }1}^{1},\text{ }\overline{q}%
_{\text{ }2}^{2},\text{ }q_{\text{ }2}^{1},\text{ }q_{\text{ }1}^{2}\}.
\end{aligned}
}

Consider now the following re-labeling of generators for the fermionic sectors $\mathfrak{f}^{(1)}$ and $\mathfrak{f}^{(3)}$, respectively, 
\EQ{
\begin{aligned}
%\mathfrak{h}_{1}&=m_{\text{ }1}^{1},\text{ \ \ \ \: }\mathfrak{h}_{2}=\overline{m}_{\text{ }%
%1}^{1},\\
T_{(++)}^{(1)} &=\overline{q}_{\text{ }2}^{1},\text{ \ \ }T_{(--)}^{(1)}=%
\overline{q}_{\text{ }1}^{2},\text{ \ \ }T_{(+-)}^{(1)}=q_{\text{ }2}^{2},%
\text{ \ }T_{(-+)}^{(1)}=q_{\text{ }1}^{1}, \\
T_{(++)}^{(3)} &=q_{\text{ }2}^{1},\text{ \ \ }T_{(--)}^{(3)}=q_{\text{ }%
1}^{2},\text{ \ \ }T_{(+-)}^{(3)}=\overline{q}_{\text{ }1}^{1},\text{ \ }%
T_{(-+)}^{(3)}=\overline{q}_{\text{ }2}^{2}.
\end{aligned}
}
They satisfy the commutation relations with the gauge algebra $\mathfrak{f}^{(0)}=\mathfrak{u}(1)\times \mathfrak{u}(1)$
\begin{equation}
\lbrack (\mathfrak{h}_{1},\mathfrak{h}_{2}),T_{(\pm \pm )}^{(a)}]=\frac{1}{2}(\pm
1,\pm 1)T_{(\pm \pm )}^{(a)},\text{ \ \ }[(\mathfrak{h}_{1},\mathfrak{h}%
_{2}),T_{(\pm \mp )}^{(a)}]=\frac{1}{2}(\pm 1,\mp 1)T_{(\pm \pm )}^{(a)},
\end{equation}%
where $a=1,3$. We have introduced the \textquotedblleft
vector\textquotedblright\ $(\mathfrak{h}_{1}=m_{\text{ }1}^{1},\mathfrak{h}_{2}=\overline{m}_{\text{ }%
1}^{1})$ in order to exhibit the gauge labels in a compact way.

\end{document}